\documentclass[]{revtex4}
\usepackage{graphicx}

\begin{document}

\title{TOPOLOGICAL SPIN CURRENT INDUCED BY NON-COMMUTING COORDINATES -AN APPLICATION TO THE  SPIN-HALL EFFECT} 
\author{D.Schmeltzer}
\affiliation{Department of Physics,\\City College of the City University of New York,\\
New York,NY,10031}
\date{\today}

\begin{abstract}

A new method for computing the spin Hall conductivity for a two dimensional electron gas in the presence of the spin Orbit interaction is presented.   The spin current is computed using the  Many-Body wave function which is  degenerated at zero momentum  The degeneracy at  $\vec{K}=0$ gives rise to non-commuting  Cartesian coordinates. 
 The non-commuting Cartesian coordinate  are a result  of an effective Aharonov-Bohm vortex at $\vec{K}=0$.  An explicit calculation for the   Rashba model is presented.  The conductivity is determined by the linear response theory  which has two parts :A-a static spin Hall conductivity which is determined by the non-commuting  coordinates and has the value  $\frac{|e|}{4\pi}$. B-a time dependent conductivity which renormalizes the static conductivity.   The value of this renormalization  depends on inelastic time scattering ,spin Orbit polarization energy and Zeeman energy.  As a result the spin Hall conductivity  vary between $\frac{|e|}{4\pi}$ and $\frac{|e|}{8\pi}$.

In the absence of a Zeeman field we find that the long time behavior is given by the renormalized conductivity $\frac{|e|}{8\pi}$ .

For relative small magnetic field the Zeeman field allows to probe continuously the spin Hall conductivity from the static  unrenormalized value  $\frac{|e|}{4\pi}$   to the fully renormalized value $\frac{|e|}{8\pi}$ . When the Zeeman energy exceeds the Fermi energy only one Fermi Dirac band is occupied and as a result the static Hall conductivity is half the static spin Hall conductivity.  

We compute the uniform magnetization without the Zeeman field and show that the spin current is covariantly conserved  and satisfies  effectively  the continuity equation.

 The effect of a time reversal scattering potential due to  a single impurity in the Rashba model  causes the the spin Hall current to decrease with the size of the system.

\end{abstract}


\maketitle

\section{INTRODUCTION}

 Two experimental groups \cite{wund,aw} have attempted to confirm the spin Hall effect  without reaching conclusive results. 
 Experimentally one observes that spins of opposite sign accumulate on opposite sides of a semiconductor in response to an external electric field. The two experimental groups \cite{wund} and \cite{aw} have  reached  different conclusions.  The Santa Barbara group \cite{aw} suggests that the spin accumulation  is extrinsic and is caused by disorder; while the Hitachi group \cite{wund} reports the experimental observation of an intrinsic  spin Hall effect  due to the   spin Orbit interaction. Recently the Santa Barbara  group \cite{Sih} have done a new experiment using a two -dimensional electron gas confined in  the (110) direction in   AlGaAs quantum wells with an electric field applied along the crystal axis. An  out of plane spin polarization is reported which might be explained by the cubic Dresselhauss   model  ruling out the disorder extrinsic effect \cite{Bern}.
   
 The proposed theories are  also controversial, a side jump disorder induced model is proposed as a mechanism  for the spin Hall effect \cite{hir}, while other  researchers have used   the spin Orbit interaction to explain the possibility for a spin Hall current, \cite{wund,sinova,halperin,sch,mol,los,culcer,yang}. It seems that the  definition of the spin current is also not clear \cite{di}. The effect of static disorder introduces vertex corrections \cite{mol} causing   the spin Hall conductivity to vanish  \cite{halperin,mol,los,sheng}. The conserved magnetic current in the presence of a orbital magnetic field has been  investigated \cite{yang,david} and a  proposal to verify  the spin Hall effect  by applying a magnetic field gradient and measuring  a charge Hall current has been suggested \cite{david}.
 
Microscopically, the spin Orbit interaction emerges from the non-relativistic limit of the Dirac equation \cite {foldy,winkler}.
At low energies the Dirac Hamiltonian for spin 1/2 electrons in an electromagnetic field is projected effectively into a two component Pauli Hamiltonian. This projection replaces the $U(1)$ gauge fields $-e \vec{A}$ and $-eA_{0}$ by $-e \vec{A}-\frac{(g-1)}{2}\mu_{B}\vec{\sigma}\times\vec{E}$ and $-eA_{0}-\frac{g}{2}\mu_{B}\vec{\sigma}\cdot\vec{B}$ where $\sigma$ is the Pauli matrix , $\frac{g}{2}\mu_{B}$ is the magnetic moment, and $\frac{-1}{2}\mu_{B}$ is the  the Thomas precession \cite{Jackson}.  $\vec{E}$ and  $\vec{B}$ are the electric and magnetic fields.
The Rashba Hamiltonian \cite{rash,winkler,mol,los} used in Solid State physics  is obtained after the replacement,$\frac{(g-1)}{2}\mu_{B}\vec{\sigma}\times\vec{E} \rightarrow \hbar k_{so}(\vec{\sigma}\times \hat{e}_{3})$ where $\hat{e}_{3}$ is the unit vector perpendicular to the two dimensional plane and  $\vec{E}$ is the  internal field  which determines the strength of the spin Orbit  momentum $k_{so}=\frac{(g-1)}{2}\mu_{B}|\vec{E}|$. For valence bands electrons the Rashba Hamiltonian is replaced by the Luttinger   Hamiltonian  \cite{lutt,winkler,mura}.

  The spin Hall conductivity has been computed  in the literature   \cite{sinova,yang} without paying attention to the singularity at zero momentum.

 We  construct an exact ground state in the presence of the spin orbit interaction which will be used to compute the spin Hall current.  As an explicit example we investigate the Rashba Hamiltonian using periodic boundary conditions with the momentum restricted to a  2-torus $T^{2}$.  We use the spinor representation to compute the commutations relations for the coordinates in the momentum representation.  This commutations  emerge from the singular spinor transformation at $K=0$ used to diagonalized the Rashba model.
   The eigenstate spinor which solves the Rashba Hamiltonian is described by a singular vector field. By a singular point we mean that the vector field is $zero $ or is not $smooth$ \cite {frankel}. In our case the spinor is multivalued at $\vec{K}=0$. The non-commutativity is a result of singular vector state at $\vec{K}=0$. In the geometrical language the non-commutativity is obtained from the knowledge of the $connection$ \cite{naka}. The $connection$ is computed in our case from the momentum derivative of the spinor-eigenstate $|K\otimes\zeta_{\alpha}(K)>$ ,($\alpha=1,2 $ represents the two component spinor which is momentum dependent ). The derivative of the $connection$ allows us to compute the geometrical $curvature$ which is interpreted as  Aharonov Bohm effect in the momentum space with a fictitious   non-zero  magnetic field at $\vec{K}=0$.
   
 The theory which emerges from the $vortex$ at $\vec{K}=0$ gives rise to the non-commuting Cartesian coordinates .
 In order to compute the spin Hall conductivity we use the linear response theory. The conductivity for a constant electric field  has two contributions: 
 
 $A$-a $static$ spin Hall conductivity of $\frac{|e|}{4\pi}$ determined by  by   zero momentum  state. 
 
 $B$- a time dependent conductivity which renormalizes the static conductivity which  is a consequence of the gapless   ground state . The time dependent corrections are given  by the $linear$ $response$ term , $\frac{-|e|}{8\pi}[1-cos(2v_{F.S.}k_{so}t)]$  where $v_{F.S.}=2\frac{(E_{F}-\epsilon_{so})}{\hbar K_{F}}$    and $\Omega_{o}= v_{F.S.} k_{so}$ are the Fermi velocity and the the spin Orbit polarization frequency.
 
 The spin Hall conductivity is given by the sum of the two parts.
  We introduce the inelastic time   scattering  $\tau_{s}$ and compare it to the   the $Fermi$ $Surface$ $Spin-Orbit$ $Polarization$ frequency $\Omega_{o}= v_{F.S.} k_{so}$.   For times $t=\tau_{s}$ which  obey $\Omega_{o} t \leq1$ we obtain that the spin Hall conductivity is given by the static vortex contribution $\frac{|e|}{4\pi}$. For $\Omega_{o} \tau_{s} >>1$  we perform a time average  and obtain that the  spin Hall conductivity takes the value, $\frac{|e|}{8\pi}$ in agreement with the value reported in the literature \cite{sinova,halperin,mol,los,culcer,yang}.  According to  ref.\cite{halperin} the spin Hall current  is sensitive to the various    scattering times . 
 For  time dependent electric fields our method allows for a simple separation between the static and  the time dependent results.
   
  The Zeeman interaction breaks the $SU(2)$ symmetry   and modifies the spin Hall conductivity   \cite {culcer}.  The presence of a $weak$  Zeeman field allows the observation  of a continuous variation of the spin Hall conductivity from the static value $\frac{|e|}{4\pi}$ (obtained for Zeeman energy  which are larger than the spin Orbit polarization energy) to the fully renormalized conductivity $\frac{|e|}{8\pi}$ (obtained for a zero magnetic field). 
  
  For extremely large  Zeeman energy which are comparable to the  the Fermi energy only one of the spin bands is populated. As a result we obtain that the static conductivity is $\frac{|e|}{8\pi}$.

-The scattering effect of a non periodic  time reversal invariant  $potential$ on the spin Hall current is investigated. We find that independent on the strength of the potential the spin Hall current vanishes with the size of the system. The scattering potential mixes the states.  The zero momentum states remain degenerated in the presence of a time reversal invariant potential. The  spectral function  weight integral  for the non zero momentum   states diverges, as a result the eigenstate  for the spin Hall current state  has a vanishing overlap with the zero momentum state causing the current to vanish.  Formally we find that the non-commuting Cartesian coordinates in the transformed basis (the basis which diagonalizes the scattering potential)  obey the same transformation rule as the geometrical curvatures \cite{frankel}. The zero momentum states remain degenerated in the presence of a time reversal invariant potential. As a result the eigenstate of the scattering potential has a vanishing overlap with the zero momentum state. Therefore for a finite system the spin Hall  current decreases with the size and $vanishes$ for an infinite system.This result is not based on statistical averages and is in qualitative agreement with the transport results obtained in the literature \cite{halperin,mol,los,sheng}. 

-The plan of this paper is as following. In chapter $II$ we introduce the Rashba Hamiltonian using the spinor representation and compute the Many-Body wave function . In chapter  $III$ we use the spinor representation to represent the  Cartesian coordinates . Using this spinor representation we compute the commutations relations. This commutations relations  emerge from the singular spinor transformation which diagonalizes the Rashba Hamiltonian. In chapter $IV$ we introduce the Heisenberg equation of motion needed for computing the  spin Hall current. We define the spin Hall  velocity  using  the Heisenberg  and Interaction picture. In chapter  $V$ we compute the Spin-Hall current. The calculation contains two parts:
 A-The $static$ spin Hall conductivity  and B-The $time$ dependent linear response spin Hall conductivity. 
In chapter $VI$ we compute the uniform magnetization induced by the electric field. We find that for $\Omega_{o} \tau_{s}\leq 1$ the uniform magnetization is finite and in the limit $\Omega_{o} \tau_{s}>>1$ it vanishes. In chapter $VII$ we show that the spin current is $covariantly$ $conserved$ and the continuity equation is satisfied to first order in the electric field.
In chapter $VIII$ we add a Zeeman interaction to the Rashba Hamiltonian and compute the spin Hall current.
In chapter $IX$ the scattering effect of  a non periodic  time reversal invariant  $potential$ on the spin Hall current is investigated. 
We have included three appendixes: In appendix A we replace the constant electric field by a time dependent vector potential. We  derive the $SU(2)$ Berry wave function  using a time dependent momentum.  In appendix B the $SU(2)$ derivation of the spin current  and the covariance conserved spin current are presented. In appendix C we compute    the charge Hall  current in the presence of both the  spin orbit interaction  and the  magnetic field .

\section{ THE MOMENTUM-SPINOR REPRESENTATION}

In this chapter we will compute the Many-Body ground state wave function for the Rashba model in the absence of an external field.
We will show that this ground state has a vortex at $\vec{K}=0$.
This wave function will be used to compute the spin Hall current.
This method is applicable to a variety of spin orbit systems  such as  the cubic  $Dresselhaus$ model  "`n"'type GaAs proposed to explain the spin Hall effect \cite{Sih,Bern}.

\textit{The Rashba model in the momentum eigenstate-spinor basis}  is given by, $\hat{h}=\hat{h}_{0}+\hat{h}_{(ext)}$  where $\hat{h}_{0}$ is the Rashba Hamiltonian.
\begin{equation}
\hat{h}_{0}=\frac{1}{2 m}[\vec{p}-\hbar{k}_{so}(\vec{\sigma}
\times\hat{e}_{3})]^{2}
\label{equation}
\end{equation}
The external potential is given by, $\hat{h}_{(ext)}= -e E^{(ext)}_{1}\cdot R^{(1)}$, where $e=-|e|$ is the electron charge, $E^{(ext)}_{1}$ is the applied electric field  at $t\geq0$ and $R^{(1)}$ is the Cartesian coordinate in the $i=1$ direction.  A careful investigation of the two component  Pauli Hamiltonian \cite{foldy} suggest that the Rashba Hamiltonian in  an external electric field  has an additional  external potential.  This is seen when  an external field is applied  to the  Pauli Hamiltonian  which is  a function  of the  total electric field, $\vec{E}=\vec{E}_{(int)}+E^{(ext)}_{1}$ . We substitute in the Pauli Hamiltonian the total electric field, $\vec{E}=\vec{E}_{(int)}+E^{(ext)}_{1}$ and find    $\frac{(g-1)}{2}\mu_{B}\vec{\sigma}\times(\vec{E}_{(int)}+E^{(ext)}_{1}) \rightarrow \hbar k_{so}(\vec{\sigma}\times \hat{e}_{3})+(\frac{\hbar k_{so}}{ E_{(int)}})(\vec{\sigma}\times E^{(ext)}_{1})$  ($E_{(int)}$ is the absolute value of the internal electric field). As a result the external potential will contain an additional  term, $\delta\hat{h}_{(ext)}=-(\frac{\hbar k_{so}}{m E_{(int)}})\sigma^{(3)} p_{2} E^{(ext)}_{1}$. This  potential is controlled by the magnetic charge  and therefore in agreement with the literature will be ignored. The fluctuations of the spin Orbit are best described by the $SU(2)$  Aharonov - Casher  spin Orbit Hamiltonian \cite{Casher} (see appendix B).
  
The Rashba Hamiltonian in eq.$1$ is given in the  momentum $\vec{p}=\hbar \vec{K}$ and coordinate $\vec{R}=i\vec{\partial}/\partial K$ representation.  We  diagonalize this Hamiltonian using a spinor representation.  The  spinor which diagonalizes the Hamiltonian is singular at $K=0$ and therefore gives rise to non commuting  coordinates.  This methodology has been used  for the Quantum Hall case where the magnetic Bloch functions have zero's in the magnetic Brillouin zone which  gives  rise  to  vortexes and  to a quantized Hall effect \cite{kohm} (see eqs.$2.13$,$3.7$,$3.8$ and $3.9$). For  a solid with a periodic potential the plane  wave functions is replaced by the Bloch functions $u_{n,K}(\vec{q})$. The transformation from the plane wave representation to the Bloch representation modifies the coordinate representation from $\vec{R}=i\vec{\partial}/\partial K$ to $\vec{r}=i\vec{\partial}/\partial K+ \vec{A}_{n,n}(\vec{K})$ where 
$\vec{A}_{n,m}(\vec{K}) = i \int \frac{d ^{2}q }{(2\pi)^{2}}u_{m,K}^{\ast} (\vec{q}) \frac{\vec{\partial}}{\partial K} u_{n,K}(\vec{q})$ \cite{zak}.
Zak has shown \cite{zak} that for a degenerated point in the Brillouin zone the Berry phase $\vec{A}_{n,n}(\vec{K})$ gives rise to a non zero band curvature, $\Omega(\vec{K},n)=\frac{\vec{\partial}}{\partial K} \times \vec{A}_{n,n}(\vec{K})$  giving rise to non-commuting coordinates, $\left[r^{(1)},r^{(2)}\right]=i\Omega(\vec{K},n)$.   
We will show that the eigenfunctions  which diagonalizes the Rashba Hamiltonian have  a singular 
 point in the momentum space and therefore the methodology used in ref. \cite{zak}  applies to our case.

The Rashba Hamiltonian  has two eigenvectors, $|K\otimes\zeta_{\alpha}(K)>$, $\alpha=1,2$ which replace the free particle-spinor $|K\otimes S>\equiv|K>|S>$ where $(|S=1>)^{T}=(1,0)$ , $(|S=2>)^{T}=(0,1)$. Due to the periodic boundary conditions the vectors $K^{(1)}$ and $K^{(2)}$ are restricted to a two dimensional $Torus$, this ensures that for a finite system  the state $\vec{K}=0$ is included in the Many-Body wave function. 
The eigenvector spinor $|K\otimes\zeta_{\alpha}(K)>$ is obtained with the help of the $SU(2)$ transformation used to diagonalize the Rashba Hamiltonian . We find, $|K\otimes\zeta_{\alpha}(K)>$=$U(\varphi(\vec{K}),\vartheta=\frac{\pi}{2})|K\otimes S>$ with the $SU(2)$ transformation which is given by,
 $U(\varphi(\vec{K}),\vartheta=\frac{\pi}{2})=\left(e^{(-i/2)\varphi(\vec{K})\sigma^{(3)}}\right)\left(e^{(-i/2)\frac{\pi}{2}\sigma^{(2)}}\right)$
  and $\varphi(\vec{K})=\arctan(\frac{- K^{(1)}}{K^{(2)}})$ is the azimuth angle in the plane, $\vec{K}=(K^{(1)},K^{(2)})$. The Rashba Hamiltonian in the diagonal  representation $ \frac{\hbar^{2}(\vec{K})^{2}}{2m}-\sigma^{(3)}\frac{\hbar^{2}k_{so}|\vec{K}|}{m}+\epsilon_{so}$ has  two eigenvalues, $\epsilon_{\sigma(\alpha)}(K))=\frac{\hbar^{2}(\vec{K})^{2}}{2m}-\sigma(\alpha)\frac{\hbar^{2}k_{so}|\vec{K}|}{m}+\epsilon_{so}\equiv\epsilon(|\vec{K}|,\sigma(\alpha))$,where $\sigma(\alpha=1)=1$ , $\sigma(\alpha=2)=-1$ and $\epsilon_{so}=\frac{\hbar^{2}k^{2}_{so}}{2 m}$ is the spin orbit single particle energy. At $\vec{K}=0$ the eigenvalues are degenerated and the spinor $|K\otimes\zeta_{\alpha}(K)>$ is $Singular $ (multivalued).  The degeneracy at  $\vec{K}=0$ gives rise to a situation where the eigenvalue function which is a function of the momentum in the  Brillouin zone returns to it's original value after a $4\pi$  rotation in the plane. Therefore  the eigenvalue function has the topology of  a connected sum of two tori $T_{g=2}$.
  
Using the eigen-spinor basis we obtain the diagonal form of the Rashba hamiltonian,
\begin{equation}
\hat{h_{0}}=\int\frac{d^{2}K}{(2\pi)^{2}}\sum_{\alpha=1}^{2}\epsilon_{\sigma(\alpha)}(|\vec{K}|)|K\otimes\zeta_{\alpha}(K)><\zeta_{\alpha}(K)\otimes K|
\label{equation}
\end{equation}

It is important to stress that the macroscopic spin current is  determined by  the ground state $Many-Body$ wave function  computed in the absence of the external field. 
  
For the Rashba model the wave function  is determined by the products of the SU(2) transformations in the momentum space which act on the   state $|G>=\prod_{\vec{K},S}|K\otimes S>$.  As a result we find that in the $transformed$ frame the $ Pauli$ $spin$ $\sigma^{(3)}$ is aligned in the plane  (the angle in the plane  is  determined by  the value of the two component momenta). 
The Many-Body ground state  $|F.S.>$ for  the Rashba model is determined by the product of the $SU(2)$ rotations.
\begin{equation}
|F.S.>=
\prod_{\vec{K},\alpha}|K\otimes\zeta_{\alpha}(K)>=\prod_{\vec{K},S}U(\varphi(\vec{K}),\vartheta=\frac{\pi}{2})|K\otimes S>
\label{ground state}
\end{equation}
The wave function $|F.S.>$ is defined in terms of the two Fermi surface  momenta ,$K_{F}^{\uparrow}$ and $K_{F}^{\downarrow}$. We introduce the notation ,   $|F.S.>\equiv|K_{F}^{\uparrow},K_{F}^{\downarrow}>$  to emphasize the fact that the occupation function is restricted by two maximal momenta $K_{F}^{\uparrow}$ and $K_{F}^{\downarrow}$.

\section{THE  CARTESIAN COORDINATES IN THE SPINOR REPRESENTATION}

The basis $|K\otimes\zeta_{\alpha}(K)>$  which builds the ground state wave function $|F.S.>$ is used to find the Cartesian coordinates $r^{(i)}$,$i=1,2$ representation ,
\begin{equation}
r^{(i)}=\int\frac{d^{2}K}{(2\pi)^{2}}\int\frac{d^{2}P}{(2\pi)^{2}}\sum_{\alpha=1}^{2}\sum_{\beta=1}^{2}r^{(i)}_{\alpha,\beta}(K,P)|K\otimes\zeta_{\alpha}(K)><\zeta_{\beta}(P)\otimes P|
\label{Hamiltonian}
\end{equation}
with the  matrix elements, 
 $r^{(i)}_{\alpha,\beta}(K,P)=I_{\alpha,\beta}\otimes\frac{id}{d K^{(i)}}\delta(\vec{K}-\vec{P})-\frac{1}{2}(\sigma^{(1)})_{\alpha,\beta}\otimes\delta(\vec{K}-\vec{P})\frac{d\varphi(\vec{K})}{d
K^{(i)}}$,
and $momentum $-$spinor$ representation;

$r^{(i)}=I\otimes\frac{i d}{d K^{(i)}}-\frac{1}{2}\sigma^{(1)}\otimes\frac{d \varphi(\vec{K})}{d K^{(i)}}\equiv R^{(i)}-\frac{1}{2}\sigma^{(1)}\otimes\frac{d \varphi(\vec{K})}{d K^{(i)}}$ where $I$ is the identity operator and 
$\sigma^{(1)}$   is the Pauli matrix. The momentum derivative $\frac{i d}{d K^{(i)}}$ represents the Cartesian coordinate $R^{(i)}$ for spinless particles and $\frac{d \varphi(\vec{K})}{d K^{(i)}}$ is the  spinor connection which is a result of the $SU(2)$  transformation which rotate  the spin frame.

The representation of the Cartesian coordinate in the momentum spinor representation given in eq.4 allows to compute the commutator for the Cartesian coordinates. In order to identify the singularity and to compute the two dimensional commutator it is advantageous to work in the complex plane, $Z=K^{(1)}+i K^{(2)}$,$\overline{Z}$,$=K^{(1)}-i K^{(2)}$. We perform a change of variables from the momentum plane $\vec{K}=(K^{(1)},K^{(2)})$ to the complex plane $Z$ and $\overline{Z}$. We find the commutator,

\begin{equation}
[r^{(1)},r^{(2)}]d K^{(1)}d K^{(2)}=\frac{\sigma^{(1)}}{4}\left[\frac{\partial}{\partial Z }(\frac{1}{\overline{Z}})d Z d \overline{Z} -\frac{\partial}{\partial\overline{Z}}(\frac{1}{Z})d \overline{Z}d Z\right]
\label{commutator}
\end{equation}
The commutator in eq.5 is zero for $\vec{K}\neq 0$. The right hand side of eq.5 represents the two dimensional delta function in the complex plane, $\delta^{(2)}(\vec{K})= \frac{1}{\pi}\frac{\partial}{\partial\overline{Z}}(\frac{1}{Z})=\frac{1}{\pi}\frac{\partial}{\partial Z }(\frac{1}{\overline{Z}})$ which is a result of the multivalued phase $\varphi(\vec{K}=0)$. At $\vec{K}=0$ the eigenvalues are degenerated and the spinor $|K\otimes\zeta_{\alpha}(K)>$ is $Singular $. The singularity in the plane at $\vec{K}=0$ represent a $vortex$ \cite{frankel} (see section 5.2 "`Closed Forms and Exact Forms ", page 156) and gives rise to non-commuting coordinates. The origin of this result follows  from the fact that the energy spectrum forms a double Torus structure in the momentum space.The central result of this section is the  result  given by the commutator in eq.$5$. This result  will be used  to compute the spin Hall conductivity in the next chapters.

\section{THE HEISENBERG EQUATION OF MOTION FOR NON-COMMUTING COORDINATES}

The singular transformation  gives rise to the following commutation relations,
$[K^{(i)},K^{(j)}]=0$, $[r^{(i)},K^{(j)}]=i\delta_{i,j}$ and 
$[r^{(i)},r^{(j)}]=(1-\delta_{i,j})\frac{-i}{2}\sigma^{(1)} 2\pi\delta^{2}(\vec{K})$ replaces the commutator,  $[R^{(i)},R^{(j)}]=0$ ($\delta_{i,j}$ represents the Kronecker delta function).
Using  the $Heisenberg$ equation of motion we find that the time derivative of the momentum $K^{(1)}$ and the coordinate $r^{(2)}$ are linear functions of the electric field $E^{ext}_{1}$; $ i\hbar\frac{d K^{(1)}(t)}{d t}=-e E^{ext}_{1}$
and,
$V_{2,H}(\vec{K},t)\equiv\frac{d r^{(2)}(t)}{d t}=(\frac{1}{i\hbar})\{[r^{(2)},\hat{h_{0}}]-e E^{ext}_{1}[r^{(2)},r^{(1)} ]\}$ is the velocity in the Heisenberg representation.
Due to the non-commutativity of the coordinates ,we find that the $velocity$ operator in the $i=2$ direction depends on  the electric field in the $i=1$ direction.  The time dependent velocity $V_{i,H}(\vec{K},t)$ (in the Heisenberg picture ) is given in terms of the Schroedinger Velocity  $V_{i}(\vec{K})$. 
 The $velocity$ and  operator in the  $Schrodinger$ picture is given by,
\begin{equation}
V_{i,H}(\vec{K},t=0)=V_{i}(\vec{K})=(\frac{d r^{(i)}(\vec{K},t)}{dt})|_{t=0}=(\frac{1}{i \hbar})([r^{(i)},\hat{h_{0}}]-e E^{ext}_{1}[r^{(i)},r^{(1)} ])|_{t=0}
\label{velocity}
\end{equation}
We will use the $first$ $quantized$ form to define the single particle operators in the momentum representation. For any single particle operator in the Schroedinger representation $O(\vec{K})$ we  define the single particle operator in the  Heisenberg $O_{H}(\vec{K},t)$ and interaction picture $O_{I}(\vec{K},t)$. 
We will apply this formulation to the  velocity and spin current.

The time dependent velocity $V_{i,H}(\vec{K},t)$ in the Heisenberg picture is given in terms of the Schroedinger Velocity   $V_{i}(\vec{K})$ defined in eq.$6$. We find, $V_{i,H}(\vec{K},t)=e^{\frac{i}{\hbar}\hat{h}(\vec{K})t)}V_{i}(\vec{K})e^{\frac{-i}{\hbar}\hat{h}(\vec{K})t}=V_{i,I}(\vec{K},t)+(\frac{-i}{\hbar})\int^{t}_{0}d t_{1}[V_{i,I}(\vec{K},t),-e r^{(1)}_{I}(\vec{K},t_{1})]E^{ext}_{1}(t_{1})+...$ where $V_{i,I}(\vec{K},t)=e^{\frac{i}{\hbar}\hat{h_{0}}(\vec{K})t}V_{i}(\vec{K})e^{\frac{-i}{\hbar}\hat{h_{0}}(\vec{K})t}$  is the velocity in the interaction picture.

 The $spin$-$velocity$ is defined according to the Noether's theorem  \cite{Weinberg} (see appendix B). We will introduce the spin velocity  in the Schroedinger picture ,
 \begin{equation}
 \ell^{(A)}_{i}(\vec{K})=\frac{1}{2}\{V_{i}(\vec{K}),\hat{\sigma}^{(A)}\}_{+}
 \label{spin-velocity}
 \end {equation}
 $\{,\}_{+}$ stands for the symmetric product and $\hat{\sigma}^{(A)}=U^{\dagger}(\vec{K})\sigma^{(A)}U (\vec{K})$ represents the transformed Pauli matrix with $A=1,2,3$. (This definition is equivalent to the symmetric projection of the $2\times2$ velocity matrix $V_{i}(\vec{K})$ into the spin space and represents the velocity of the particle with a given spin polarization.)
 
 The spin-velocity in the $i=2$ direction and polarization $A=3$  takes in the  Heisenberg representation the form,
\begin{equation} \ell^{(3)}_{i,H}(\vec{K},t)=\ell^{(3)}_{i,I}(\vec{K},t)+(\frac{-i}{\hbar})\int^{t}_{0}d t_{1}[\ell^{(3)}_{i,I}(\vec{K},t),-e r^{(1)}_{I}(\vec{K},t_{1})]E^{ext}_{1}(t_{1})+...
 \label{interaction picture}
 \end{equation}
 where  $\ell^{(3)}_{2,I}(\vec{K},t)$ and $r^{(1)}_{I}(\vec{K},t)$  are  defined according to the $Interaction$ picture,
\begin{equation} \ell^{(3)}_{2,I}(\vec{K},t)=e^{\frac{i}{\hbar}\hat{h_{0}}(\vec{K})t}\ell^{(3)}_{2}(\vec{K})e^{\frac{-i}{\hbar}\hat{h_{0}}(\vec{K})t}\equiv\ell^{(3,0)}_{2,I}(\vec{K},t)+\ell^{(3,ext.)}_{2,I}(\vec{K},t)
\label{picture}
\end{equation}

 Where $\ell^{(3,0)}_{2,I}(\vec{K},t)=e^{\frac{i}{\hbar}\hat{h_{0}}(\vec{K})t}\ell^{(3,0)}_{2}(\vec{K})e^{\frac{-i}{\hbar}\hat{h_{0}}(\vec{K})t}$ represents the $homogeneous$ spin velocity   in the Schroedinger picture  ,$\ell^{(3,0)}_{2}(\vec{K})=\frac{1}{2}\{ \frac{1}{i\hbar}([r^{(2)},\hat{h_{0}}])|_{t=0},\hat{\sigma}^{(3)}\}_{+}$.  The second term in eq.$9$ represent the static   spin-velocity generated by the $non$-$commuting$ Cartesian. This term is generated by  external electric field $E^{ext}_{1}$ at $t=0$  (the second term in eq.$6$).  We find that the  external spin velocity $\ell^{  (3,ext.)}_{2}(\vec{K})$ in the Interaction picture is the same as the Schroedinger picture $\ell^{  (3,ext.)}_{2}(\vec{K})=\ell^{(3,ext.)}_{2,I}(\vec{K},t)$ 
 where ,$\ell^{(3,ext.)}_{2}(\vec{K})=\frac{1}{2}\{ \frac{1}{i\hbar}(-e E^{ext}_{1}[r^{(2)},r^{(1)} ])|_{t=0},\hat{\sigma}^{(3)}\}_{+}$.

The linear response result in our case is given by  eq.$8$:

 The first term will give rise to the $static$  linear response part , $\ell^{  (3,ext.)}_{2}(\vec{K})=\ell^{(3,ext.)}_{2,I}(\vec{K},t)=\frac{1}{2}\{ \frac{1}{i\hbar}(-e E^{ext}_{1}[r^{(2)},r^{(1)} ])|_{t=0},\hat{\sigma}^{(3)}\}_{+}$.
 This part is due to the non-commuting coordinates. 
 
 The second  term  in eq.$8$  (the term which is linear in the external electric field  $E^{ext}_{1}$ ) will give rise to the $time$ $dependent$ linear response when we substitute in the commutator of eq.$8$ the homogeneous 
spin velocity operator $\ell^{(3,0)}_{2,I}(\vec{K},t)$ .

We will  compute the expectation values using the second quantized form .  We introduce  the two component Spinors in the momentum representation $\Psi^{\dagger}(\vec{K})$ and  $\Psi(\vec{K})$ 
which act on the Many-Body ground state,$|F.S.>$ . We introduce the Heisenberg operator in the second quantized form , $\hat{O}_{H}=\int \frac{d^{2}K}{(2\pi)^{2}}\Psi^{\dagger}(\vec{K})O_{H}(\vec{K},t)\Psi^{\dagger}(\vec{K})$ .
 The second quantized form  of the $spin$ current $J^{(3)}_{2,H}$ in the Heisenberg picture is obtained from the Heisenberg representation of  the $single$  $particle$ operators. The spin current is obtained in the limit $q\rightarrow0$ by taking the expectation value of the $spin$ current operator with respect the ground state $|F.S.>$ .

\begin{eqnarray}
J^{(3)}_{2}&=& <F.S.|J^{(3)}_{2,H}|F.S.>\nonumber\\&=& lim_{\vec{q}\rightarrow0}(\frac{\hbar}{2})\int \frac{d^{2}K}{(2\pi)^{2}}<F.S.|\Psi^{\dagger}(\vec{K})(\ell^{(3)}_{2,H}(\vec{K},t)e^{i\vec{q}\cdot\vec{r}_{H}(\vec{K},t)})\Psi(\vec{K})|F.S.>\nonumber\\& =& J^{(3,ext.-static)}_{2}+\delta J^{(3,time-dependent)}_{2}
\label{heisenberg current} 
\end{eqnarray}
$\vec{r}_{H}(\vec{K},t)$ is the coordinate operator in the Heisenberg picture and the exponential acts as a shift operator in the momentum space, $e^{i\vec{q}\cdot\vec{r}(\vec{K})}\Psi(\vec{K})=\Psi(\vec{K}-\vec{q})$.

The first term $J^{(3,ext.-static)}_{2}$ represents the $static$ linear response  current due to the  the vortex at $\vec{K}=0$.
The second term 
$\delta J^{(3,time-dependent)}_{2}$ is the $time$ $dependent$  linear response . This term gives rise to renormalization effect of the static current. In the presence of a gap this renormalization effect are negligible. (This is the situation for the integer Quantum Hall where the conductivity is given by the $static$ part.)

\section{THE  SPIN-HALL CURRENT}

In order to compute the spin Hall current we will use the ground state wave function $|F.S>$ characterized by the $Fermi-Dirac$ occupation function $f_{F.D.}[\epsilon(|\vec{K}|,\sigma(\alpha))-E_{F}]$ which at $T=0$ is given by the $step$ function,$\theta[\epsilon(|\vec{K}|,\sigma(\alpha))-E_{F}]$.
 The spin Hall current has two parts ;the $static$ linear response   given by the first term in eq.$8$   and the $time$ $dependent$ linear response term given by the second term in eq.$8$.
 
A-The $static$ linear response current $J^{(3,ext.-static)}_{2}$.

 The formula for the static spin current in the $i=2$ direction is given by the term, $J^{(3,ext.-static)}_{2}$. The expectation value with respect the ground state $|F.S.>$  at zero temperature gives,

\begin{eqnarray}
J^{(3,ext.-static)}_{2}&=&(\frac{\hbar}{2})\int \frac{d^{2}K}{(2\pi)^{2}}<F.S.|\Psi^{\dagger}(\vec{K})[\ell^{(3,ext.)}_{2}(\vec{K})]\Psi(\vec{K})|F.S.>\nonumber\\&=&
\frac{1}{(2\pi)^{2}}\sum_{\alpha=1}^{2}\int\int \frac{-e}{2i}E^{(ext)}_{1} \sigma^{(1)}[r^{(2)},r^{(1)} ] \theta[\epsilon(|\vec{K}|,\sigma(\alpha))-E_{F}]d K^{(1)}d K^{(2)}\nonumber\\&=&\frac{1}{(2\pi)^{2}}\sum_{\alpha=1}^{2}\int\int \frac{e}{2i}E^{(ext)}_{1}\theta[\epsilon(|\vec{K}|,\sigma(\alpha))-E_{F}]\frac{( \sigma^{(1)})^{2}}{4}\left[\partial_{Z}(\frac{1}{\overline{Z}})d Z d \overline{Z} -\partial_{\overline{Z}}(\frac{1}{Z})d \overline{Z}d Z\right]\nonumber\\& =&\frac{e}{2}E^{(ext)}_{1}\frac{1}{2}\frac{\pi}{(2\pi)^{2}}\sum_{\alpha=1}^{2}[\frac{1}{2\pi i}\oint \frac{d\overline {Z}}{\overline{Z}}-\frac{1}{2\pi i}\oint \frac{d Z}{Z}]=\frac{-e}{4\pi}E^{(ext)}_{1}
\label{spin current}
\end {eqnarray}

The second row in eq.$11$ is a function of the commutator  $[r^{(1)},r^{(2)}]$. The third row in eq.$11$ is obtained after we replace the commutator $[r^{(1)},r^{(2)}]$ with his complex representation given in eq.5 . In the third  row we recognize the two dimensional delta function ,$\delta^{(2)}(\vec{K})= \frac{1}{\pi}\frac{\partial}{\partial\overline{Z}}(\frac{1}{Z})=\frac{1}{\pi}\frac{\partial}{\partial Z }(\frac{1}{\overline{Z}})$ which enable us to replace a two dimensional integral with a $closed$ line integral in agreement with the $Stokes$ theorem. The theorem allows us to replace the two dimensional integral over the full Fermi see with  a closed line integral  at the boundary of  on  the Fermi surface (the two Fermi surfaces, for spin up and spin down) . As a result we find that the $static$ $Spin-Hall$ current in eq.$11$ is replaced by  a $Fermi -Surface$ (the line integral in the fourth row) integral  gives the $exact$ value of $\frac{-e}{4\pi}=\frac{|e|}{4\pi}$ for the  $static$ $Spin-Hall$ $conductivity$. The origin of this $exact$ result,  is caused by the current which is carried  by the vortex at $K=0$. The Fermi Dirac occupation function at $K=0$ takes the value of one for spin up and  spin down, therefore the contribution from the two spin polarization in eq.$11$ are equal.

B-The linear response $time$ $dependent$ linear response spin Hall current $\delta J^{(3,time-dependent)}_{2}$

The linear response ,$time$ $dependent$  spin Hall current is given by the second term in eq.9. We use the interaction picture  for  the spin velocity ,Cartesian coordinate and the Pauli matrix . $\ell^{(3,0)}_{2,I}(\vec{K},t)=e^{\frac{i}{\hbar}\hat{h_{0}}(\vec{K})t}\ell^{(3,0)}_{2}(\vec{K})e^{\frac{-i}{\hbar}\hat{h_{0}}(\vec{K})t}$,  $r^{(1)}_{I}(\vec{K},t)=e^{\frac{i}{\hbar}\hat{h_{0}}(\vec{K})t}r^{(1)}_{I}(\vec{K})e^{\frac{-i}{\hbar}\hat{h_{0}}(\vec{K})t}$, $\sigma^{(1)}_{I}(\vec{K},t_{1})=e^{\frac{i}{\hbar}\hat{h_{0}}(\vec{K})t}\sigma^{(1)}e^{\frac{-i}{\hbar}\hat{h_{0}}(\vec{K})t}$.
 \begin{eqnarray}\nonumber
Tr[\ell^{(3,0)}_{2,I}(\vec{K},t),(-e)r^{(1)}_{I}(\vec{K},t)]&=& Tr[(\frac{\hbar}{m})K^{(2)}\hat{\sigma}^{(3)}_{I}(\vec{K},t),\frac{e}{2}\sigma^{(1)}_{I}(\vec{K},t_{1})\frac{d\varphi}{d K^{(1)}}]\nonumber\\&=&
\frac{-e }{2}(\frac{\hbar}{m})K^{(2)}\frac{d\varphi}{d K^{(1)}}Tr[\sigma^{(1)}_{I}(\vec{K},t),\sigma^{(1)}_{I}(\vec{K},t_{1})] \label{trace}
\end{eqnarray}
$"'Tr"'$ represents the trace over the spin space . Using the interaction picture of the  spin operators we evaluate the time dependent commutator,
\begin{equation}
[\sigma^{(1)}_{I}(\vec{K},t),\sigma^{(1)}_{I}(\vec{K},t_{1})]=(-2i)\sigma^{(3)}\sin[(\frac{2\hbar k_{so}|\vec{K}|}{m})(t-t_{1})]
\label{sigma}
\end{equation}
From equations $12$and $13$  we obtain the linear response time dependent term  $\delta J^{(3,time-dependent)}_{2}$ which is computed from the second term in equation $8$ by substituting $\ell^{(3)}_{2,I}(\vec{K},t)$  with $\ell^{(3,0)}_{2,I}(\vec{K},t)$.

\begin{eqnarray}\nonumber
& &\delta J^{(3,time-dependent)}_{2}=(\frac{-i}{\hbar})(\frac{\hbar}{2})Tr\int\frac{d^{2}K}{(2\pi)^{2}}\int^{t}_{0}d t_{1}[\ell^{(3,0)}_{2,I}(\vec{K},t),-e r^{(1)}_{I}(\vec{K},t_{1})]E^{ext}_{1}(t_{1})\nonumber\\&=& \frac{e}{8\pi}\int^{\infty}_{0} d(\frac{\hbar^{2}K^{2}}{2 m})[f_{F.D.}(\frac{\hbar^{2}K^{2}}{2m}+\epsilon_{so}-\frac{\hbar^{2}k_{so}}{m}|\vec{K}|-E_{F})-f_{F.D.}(\frac{\hbar^{2}K^{2}}{2m}+\epsilon_{so}+\frac{\hbar^{2}k_{so}}{m}|\vec{K}|-E_{F})]\nonumber \\& &\cdot\int^{t}_{0}d(-t_{1})\sin[(\frac{2\hbar k_{so}|\vec{K}|}{m})(t-t_{1})]E^{ext}_{1}(t_{1})=\frac{e}{8\pi}\int^{t}_{0} d(-2\Omega_{o} t_{1})\sin[( 2\Omega_{o}(t-t_{1})]E^{ext}_{1}(t_{1})
\label{integration}
\end{eqnarray}

At zero temperature the  difference between the two Fermi - Dirac step functions for spin up and spin down  is equal  to  the Fermi-Surface Spin-Orbit Polarization energy  $\hbar\Omega_{o}$ multiplied by the delta function $\delta[(\frac{\hbar^{2}K^{2}}{2m}+\epsilon_{so}-E_{F})]$.  The polarization frequency $\Omega_{o}= v_{F.S.} k_{so}$ is defined in term of the spin Orbit momentum   $k_{so}=\frac{(g-1)}{2}\mu_{B}|\vec{E}|$  and the Fermi velocity $v_{F.S.}=2\frac{(E_{F}-\epsilon_{so})}{\hbar K_{F}}$. 
 The momentum integration is replaced by the energy integration  $d(\frac{\hbar^{2}K^{2}}{2m})$ restricted by the delta function $\delta[(\frac{\hbar^{2}K^{2}}{2m}+\epsilon_{so}-E_{F})]$ . As a result  we find that the $linear$ $response$  time dependent current for a $time$ $dependent$ electric field is given by  $\delta J^{(3,time-dependent)}_{2}=\frac{e}{8\pi}\int^{t}_{0} d(-2\Omega_{o} t_{1})\sin[( 2\Omega_{o}(t-t_{1})]E^{ext}_{1}(t_{1})$ (see eq.$14$)
 For the special case where the electric field is $time$ $independent$  the time integration in eq. $14$ gives the linear response current,
  $\delta J^{(3,time-dependent)}_{2}=\frac{e}{8\pi}[1-\cos( 2\Omega_{o} t)]E^{ext}_{1}$.
 
Combining the two results ,the one obtained in equation $11$  with the one given  in  equation  $14$ we obtain   the spin Hall current for the time independent  electric field $E^{ext}_{1}$.
\begin{equation}
J^{(3)}_{2}\equiv J^{(3,ext.-static)}_{2}+\delta J^{(3,time-dependent)}_{2} =\frac{-e}{4\pi}E^{ext}_{1}+\frac{e}{8\pi}[1-\cos( 2\Omega_{o} t)]E^{ext}_{1}
\label{sum}
\end{equation}
The result in equation $14$ can be considered for different limits  of  inelastic scattering  (the inelastic scattering time   describes the temperature or other incoherent scattering such as  spin waves or  spin exchange)  . The inelastic scattering is described by the time $\tau_{s}$  which has to be compared with the spin Orbit polarization frequency  $\Omega_{o}$. In the limit $\Omega_{o}\tau_{s}\leq1$ the second term in eq.$14$ can be ignored and one finds that the spin Hall conductivity is $\frac{|e|}{4\pi}$.
In the opposite limit $\Omega_{o}\tau_{s}>>1$ the second term can be replaced by a time average and we find that the spin Hall conductivity is given by $\frac{|e|}{8\pi}$ in agreement with the result obtained in the literature \cite{sinova,halperin,yang,culcer}.
.   
 
\section{THE MAGNETIZATION}

The application of an electric field in the $i=1$ direction  generates an uniform magnetization with an in plane spin polarization $M^{(A=2)}$  (in the $i=2$) direction which is $time$ dependent and vanishes in the long time limit. 
The uniform magnetization is given by , 
\begin{equation}
M^{(A)}(t)=\frac{ \hbar}{2}\int\frac{d^{2}K}{(2\pi)^{2}}<F.S.|\Psi^{\dagger}(\vec{K})
\hat\sigma^{(A)}_{H}(\vec{K},t)\Psi(\vec{K})|F.S.>
\label{magnetization}
\end{equation}
As a result of the $SU(2)$ transformation the spin operators are transformed; $\hat\sigma^{(1)}(\vec{K})=\sigma^{(3)}\cos\varphi(\vec{K})-\sigma^{(2)}\sin\varphi(\vec{K})$, $\hat\sigma^{(2)}(\vec{K})=\sigma^{(2)}\cos\varphi(\vec{K})-\sigma^{(3)}\sin\varphi(\vec{K})$ and $\hat\sigma^{(3)}=-\sigma^{(1)}$. The Pauli matrices $A=1,2,3$  in the Heisenberg representation are represented in terms of the  Pauli  matrices and  the coordinate $r_{I}^{(1)}$ in the Interaction picture.
\begin{equation}
\sigma^{(A)}_{H}(\vec{K},t)=\sigma^{(A)}_{I}(\vec{K},t)-\frac{i}{\hbar}\int^{t}_{0}d t_{1}[\sigma^{(A)}_{I}(\vec{K},t),-e r_{I}^{(1)}(\vec{K},t_{1})]E^{ext}_{1}(t_{1})...
\label{spindynamics}
\end{equation}
Using the explicit form of the Pauli matrices and the $SU(2)$ angular $\varphi(\vec{K})$ dependence,
$\sin\varphi(\vec{K})=-\frac{K^{(1)}}{|\vec{K}|}$,and $\cos\varphi(\vec{K})=\frac{K^{(2)}}{|\vec{K}|}$ we find  that
the  only non zero term is $M^{(2)}$. This is the  magnetization with the $A=2$ polarization  (this is the only term which is   invariant under the transformation $\vec{K}\rightarrow -\vec{K}$  and has  a finite  trace over the Pauli matrices ).
We have,

\begin{eqnarray}
 M^{(2)}(t)&=&\frac{ \hbar}{2}\int\frac{d^{2}K}{(2\pi)^{2}}<F.S.|\Psi^{\dagger}(\vec{K})\{\cos\varphi(\vec{K})[\sigma^{(2)}_{I}(\vec{K},t)-\frac{i}{\hbar}\int^{t}_{0}d t_{1}[\sigma^{(2)}_{I}(\vec{K},t),-e r_{I}^{(1)}(\vec{K},t_{1})]E^{ext}_{1}(t_{1})]\nonumber\\&-&\sin\varphi(\vec{K})[\sigma^{(3)}_{I}(\vec{K},t)-\frac{i}{\hbar}\int^{t}_{0}d t_{1}[\sigma^{(3)}_{I}(\vec{K},t),-e r_{I}^{(1)}(\vec{K},t_{1})]E^{ext}_{1}(t_{1})]\}\Psi(\vec{K})|F.S.>
\label{mag}
\end{eqnarray}

The commutator in the last equation is proportional to $\frac{d \varphi(\vec{K})}{d K^{(1)}}=-\frac{K^{(2)}}{|\vec{K}|^{2}}$ therefore only the product with the term proportional to  $\cos\varphi(\vec{K})=\frac{K^{(2)}}{|\vec{K}|}$  will be nonzero.
As a result we obtain,
\begin{eqnarray}
 M^{(2)}(t)&=&(\frac{ \hbar}{2})(\frac{-i}{\hbar})(\frac{-e}{2})\int\frac{d^{2}K}{(2\pi)^{2}}\nonumber\\& &<F.S.|\Psi^{\dagger}(\vec{K})\{\cos\varphi(\vec{K})\frac{d \varphi(\vec{K})}{d K^{(1)}}\int^{t}_{0}d t_{1}[\sigma^{(2)}_{I}(\vec{K},t), \sigma^{(2)}_{I}(\vec{K},t_{1})]E^{ext}_{1}(t_{1})\}\Psi(\vec{K})|F.S.>\nonumber\\&=&\frac{-e}{8\pi}\int ^{\infty}_{0}d K[f _{F.D.}(\frac{\hbar^{2}K^{2}}{2m}+\epsilon_{so}-\frac{\hbar^{2}k_{so}}{m}|\vec{K}|-E_{F})-f_{F.D.}(\frac{\hbar^{2}K^{2}}{2m}+\epsilon_{so}+\frac{\hbar^{2}k_{so}}{m}|\vec{K}|-E_{F})]\nonumber\\& &\int^{t}_{0}d t_{1}\cos(\frac{\hbar 2 k_{so}}{m}|\vec{K}|(t-t_{1})E^{ext}_{1}(t_{1}) \nonumber\\&=&\frac{-e\hbar}{8\pi}\int^{\infty}_{0} d K(\frac{2\hbar  k_{so} K}{m})\delta(\frac{\hbar^{2}K^{2}}{2m}+\epsilon_{so}-E_{F})\int^{t}_{0}d t_{1}\cos(\frac{2\hbar k_{so}K}{m}(t-t_{1}))E^{ext}_{1}(t_{1})
\label{mag2}
\end{eqnarray}
For a constant electric field we perform the time integral and find  the time dependent magnetization.
\begin{equation}
M^{(2)}(t)=(\frac{\hbar}{2})(\frac{1}{4\pi})k_{so}[\frac{-e E^{ext}_{1}t}{\hbar}][\frac{\sin(2\Omega_{o}t)}{2\Omega_{o}t}]  
\label{static}
\end{equation}
where $\Omega_{o}$ is the the polarization frequency $\Omega_{o}= v_{F.S.} k_{so}$  defined in the previous chapter. We observe that for inelastic scattering times $\tau_{s}$ which obey $\Omega_{o}\tau_{s}\leq 1$ the magnetization is $finite$,
 $M^{(2)}\approx(\frac{\hbar}{2})(\frac{1}{4\pi})k_{so}[\frac{-e E^{ext}_{1}\tau_{s}}{\hbar}]$.  In the opposite limit  we perform the  time average and find that the uniform  magnetization  $vanishes$.

\section{THE COVARIANTLY CONSERVED SPIN CURRENT}

Following  Noether's theorem \cite{Weinberg,Frohlich}  (see Appendix B) we compute    the spin currents . 
 We apply this formulation  to the $Rashba$ model and find from  eq.$B9$ that the spin current is \textit{covariantly conserved}.
  The three spin polarizations currents obey in the long wave limit limit $q \rightarrow 0$  the following  continuity equation ;
 $\partial_t J_0^{(1)} + i \vec{q} \cdot \vec{J}^{(1)}(\vec{q}) = 0$, $\partial_t J_0^{(2)} + i \vec{q} \cdot \vec{J}^{(2)}(q) = 0$. Only the $A=3$ component \textit{violates the continuity equation}
\begin{eqnarray}
& &\partial_t J_0^{(3)} + i \vec{q} \cdot \vec{J}^{(3)}(\vec{q}) =  2 k_{so} (J_1^{(1)}(q) + J_2^{(2)}(q))\nonumber\\&_{\overrightarrow{q \rightarrow0}}& 2 k_{so}(\frac{\hbar}{2}) <F.S.| \int \frac{d^2 K}{(2\pi)^{2}} \Psi^+(\vec{K},t) [ \frac{1}{2}\{{V}_{1,H}(\vec{K},t),\hat{ \sigma_{H}}^{(1)}(\vec{K},t)\}_{+} + \frac{1}{2}\{{V}_{2,H}(\vec{K},t),\hat{ \sigma}_{H}^{(2)}(\vec{K},t)\}_{+} ] \Psi(\vec{K},t)|F.S.>\nonumber\\&=&2 k_{so}(\frac{\hbar}{2}) \int \frac{d^2 K}{(2\pi)^{2}}<F.S.|\Psi^+(\vec{K})[\frac{1}{2}\{\sigma_{H}^{(3)}(\vec{K},t),(\vec{V}_{H}(\vec{K},t)\times \frac{\vec{K}}{|\vec{K}|})\}_{+}\nonumber\\&+&\frac{1}{2}\{\sigma_{H}^{(2)}(\vec{K},t),(\vec{V}_{H}(\vec{K},t)\cdot \frac{\vec{K}}{|\vec{K}|})\}_{+}]\Psi(\vec{K})|F.S.>
\label{covariant}
\end{eqnarray}
This  equation has been obtained from the Heisenberg representation of the spin velocity  and the  $SU(2)$  transformed  Pauli matrices.( $\hat{ \sigma}^{(1)}= \sigma^{(3)}\cos \varphi(\vec{K}) -\sigma^{(2)}\sin \varphi(\vec{K})$,$\hat{ \sigma}^{(2)}= \sigma^{(2)}\cos \varphi(\vec{K})) -\sigma^{(3)}\sin \varphi(\vec{K})$). The right hand side of equation $21$ is evaluated using the Heisenberg representation of the operators.The velocity and spin operators are a function of the external electric field, see eqs.$8$and $17$. To first order in the electric field we observed that the right hand side term vanishes.(The first term on the right hand side of  the equation  vanishes  as a result  of to the momentum  integration .and the second term vanishes as a result of the expectation value with respect the Pauli matrices.) Therefore  we conclude that the  $continuity$ $equation$ is  effectively satisfied.

\section{THE SPIN HALL EFFECT IN THE PRESENCE OF A ZEEMAN INTERACTION}

 From the analysis in chapters $IV$ and $V$ we learn that the static spin current is determined by the state $\vec{K}=0$ . Since the Fermi-Dirac  step function for the state  $\vec{K}=0$ at zero temperature  is  not  affected by a Zeeman field  (if this energy is much less than the Fermi energy) the static spin-Hall current will be same as for the zero magnetic field (see eq.$11$). The effect of the magnetic field will be to  renormalize the  static current.
 
 We will present first the modification caused by the Zeeman field to the Rashba model. The  Zeeman energy $b_{3}=\frac{1}{2}g\mu_{B}B$ ( $B$ is  the effective  magnetic field) changes the the single  particle eigenvalues from ,  $\epsilon_{\sigma(\alpha)}(K))=\frac{\hbar^{2}(\vec{K})^{2}}{2m}-\sigma(\alpha)\frac{\hbar^{2}k_{so}|\vec{K}|}{m}+\epsilon_{so}$  to  $\epsilon_{\sigma(\alpha)}(K))=\frac{\hbar^{2}(\vec{K})^{2}}{2m}-\sigma(\alpha)\Delta(K)+\epsilon_{so}$ where $\Delta(K)=\sqrt{b^{2}_{3}+(\frac{\hbar^{2}k_{so}|\vec{K}|}{m})^{2}}$.     The $SU(2)$ transformation  is replaced by  ,
 $U(\varphi(\vec{K}),\vartheta(\vec{K}))=e^{(-i/2)\varphi(\vec{K})\sigma^{(3)}}e^{(-i/2)\vartheta(\vec{K})\sigma^{(2)}}$ with      the polar angle $\vartheta=\frac{\pi}{2}$  (zero magnetic field) replaced by $\tan\vartheta(K)=\frac{\hbar^{2}k_{so}|\vec{K}|}{m b_{3}}$ (finite magnetic field).
 
 Following the steps described in equations $3-6$ we find that  the   polar angle $\vartheta(\vec{K})$  modifies the transformed  the Cartesian coordinates $i=1,2$.
 
\begin{equation}
r^{(i)}(\vec{K})=R^{(i)}(\vec{K})-\frac{1}{2}[\sigma^{(1)}\sin \vartheta(\vec{K})\frac{d\varphi(\vec{K})}{d K^{(i)}}-\sigma^{(2)}\frac{d\vartheta(\vec{K})}{d K^{(i)}}-\sigma^{(3)}\cos \vartheta(\vec{K})\frac{d\varphi(\vec{K})}{d K^{(i)}}]
\label{gauge}
\end{equation}

 and the commutation relations are transformed , 
 
\begin{equation}
\left[r^{(1)},r^{(2)}\right]=\frac{i}{2}\left[-\sigma^{(1)}\sin\vartheta(\vec{K})+\sigma^{(3)}\cos\vartheta(\vec{K})\right]2\pi\delta^{2}(K)
\label{zcommutator}
\end{equation}

 where the  transformed Pauli matrix is given by, $\sigma^{(3)}\rightarrow U^\dag \sigma^{(3) }U = \sigma^{(3)} \cos \vartheta(\vec{K}) - \sigma^{(1)} \sin \vartheta(\vec{K})$   and the delta function ,$\delta^{(2)}(\vec{K})= \frac{1}{\pi}\frac{\partial}{\partial\overline{Z}}(\frac{1}{Z})=\frac{1}{\pi}\frac{\partial}{\partial Z }(\frac{1}{\overline{Z}})$ 
  
 At zero temperature the ground state is described by the two $Fermi-Dirac$  for spin up and spin down)step functions, $\theta[\epsilon(|\vec{K}|,\sigma(\alpha))-E_{F}]$. 
 We will consider two cases:

 a) The magnetic Zeeman energy is much less than the Fermi energy  therefore the  sate $\vec{K}=0$ with spin up or spin down have equal occupation, $\theta[\epsilon_{so}+b_{3}-E_{F}]=\theta[-b_{3}+\epsilon_{so}-E_{F}]=1$. For this case the static spin Hall current  is identical to the case without the Zeeman field .Following the method used in section $V$ we find that the spin Hall current  is given by two contributions ;the static one (see eq.$11$) and the the time dependent contribution (see eq.14).
The sum of the two parts is given by an expression similar to the one given in eq.$15$. 

\begin{equation} J^{(3)}_{2}=\frac{-e}{4\pi}E^{(ext)}_{1}+\frac{e}{8\pi}(\frac{(\hbar\Omega_{o})^{2}}{(\hbar\Omega_{o})^{2}+b_{3}^{2}})[1-\cos( 2(\Omega^{2}_{o}+(\frac{b_{3}}{\hbar})^{2})^{\frac{1}{2}}t)]E^{ext}_{1}
\label{spin zcurrent}
\end {equation}

 The static spin Hall conductivity is determined by the degenerate state at $\vec{K}=0$
and is given by the first term in eq.$24$ which is identical with the first term in eq.$15$. The second term in eq.$24$ represents the time dependent linear response part. Here we observe that the Zeeman magnetic interaction controls  the renormalization effects.
 The  Zeeman interaction modifies the value of the commutator given in eq.$12$.  which is replaced by the term $\sin^{2}\vartheta(\vec{K})= (\frac{(\hbar\Omega_{o})^{2}}{(\hbar\Omega_{o})^{2}+b_{3}^{2}})$.  
 When the   Zeeman energies $b_{3}$  is  larger  than polarization energy $\hbar\Omega_{o}$ the linear response term can be ignored and the spin Hall current is determined by the $static$   solution, $J^{(3)}_{2}\approx\frac{|e|}{4\pi}E^{(ext)}_{1}$.
For Zeeman energies which are comparable or less than the polarization energy  we find in the long time limit, $(\Omega^{2}_{o}+(\frac{b_{3}}{\hbar})^{2})^{\frac{1}{2}}\tau_{s}>>1$   that the spin Hall current is given by,
 $\frac{|e|}{4\pi}[1-\frac{1}{2}(\frac{(\hbar\Omega_{o})^{2}}{(\hbar\Omega_{o})^{2}+b_{3}^{2}})]$
 
Therefore in this case the magnetic field can be used to vary continuously  the spin Hall conductivity from the $static$ conductivity $\frac{|e|}{4\pi}$ for $\frac{b_{3}}{\hbar\Omega_{o}}>>1$  ( for large magnetic fields comparable to the polarization energy ) to the    fully $renormalized$ spin Hall conductivity  $\frac{|e|}{8\pi}$  for $b_{3}=0$ (zero magnetic).

b)-When the Zeeman interaction is comparable or larger than the  Fermi energy only one of the Dirac step function contributes. For this case  we find a spin Hall conductivity is determined by the $static$ part (without normalization by the time dependent linear response term).In this case   only $one$ of the Dirac step functions contribute to the conductivity ,$\frac{|e|}{8\pi}$.

\section{THE SPIN-HALL EFEECT IN THE PRESENCE OF A NON-PERIODIC SCATTERING POTENTIAL}
We will investigate the effect of a non-periodic time reversal invariant potential $V(\vec{r})$ on the Rashba hamiltonian in the absence of inelastic and scattering and Zeeman field. We will see that the spin Hall current depends on the ratio $\frac{L}{\lambda_{so}}$ where $L$ is the size  of the system  and $\lambda_{so}$is the spin orbit length defined by the inverse of the spin Orbit momentum $\lambda_{so}=\frac{2\pi}{k_{so}}$ where $k_{so}=\frac{\omega_{o}}{v_{F.S.}}=\frac{g-1}{2}\mu_{B}|\vec{E}|$.

 The hamiltonian, $h= h_{0}+V(\vec{r})$ is represented with the help of the the Rashba eigenstates basis,
\begin{eqnarray}
h&=&\int\frac{d^{2}K}{(2\pi)^{2}}\sum_{\alpha=1}^{2}\epsilon_{\sigma(\alpha)}(|\vec{K}|)|K\otimes\zeta_{\alpha}(K)><\zeta_{\alpha}(K)\otimes K|\nonumber\\ & &+\int\frac{d^{2}K}{(2\pi)^{2}}\int\frac{d^{2}P}{(2\pi)^{2}}\sum_{\alpha=1}^{2}\sum_{\beta=1}^{2}\hat{V}_{\alpha,\beta}(\vec{K}-\vec{P})|K\otimes\zeta_{\alpha}(K)><\zeta_{\beta}(P)\otimes P|
\label{scattering }
\end{eqnarray}
Where the matrix elements of the scattering potential are given by the product of the Fourier component in the momentum representation,$V(\vec{K}-\vec{P})$ with the $SU(2)$ matrix elements,      $U(\varphi(\vec{K}),\vartheta=\frac{\pi}{2})=\left(e^{(-i/2)\varphi(\vec{K})\sigma_{3}}\right)\left(e^{(-i/2)\frac{\pi}{2}\sigma_{2}}\right)$. We define the transformed potential, $\hat{V}_{\alpha,\beta}(\vec{K}-\vec{P})=\sum_{\lambda=1}^{2}U^{\dagger}(\varphi(\vec{K}),\vartheta=\frac{\pi}{2})_{\alpha,\lambda}V(\vec{K}-\vec{P})U(\varphi(\vec{P}),\vartheta=\frac{\pi}{2})_{\lambda,\beta}$.
 The $eigenstates$ $|\Phi^{(\alpha)}(K)>$ of the hamiltonian, $h= h_{0}+V(\vec{r})$ replaces the eigenstate $|K\otimes\zeta_{\alpha}(K)>$ given in eq.2.
\begin{eqnarray}
|\Phi^{(\alpha)}(K)>&=&\frac{1}{\sqrt{N^{(\alpha)}(K)}}[|K\otimes\zeta_{\alpha}(K)>+(1-\delta_{\vec{K},0})\sum_{\lambda=1}^{2}(\delta_{\alpha,1}\delta_{\lambda,2}+\delta_{\alpha,2}\delta_{\lambda,1})\mathcal A^{(\alpha)}_{K}(K,\lambda)|K\otimes\zeta_{\lambda}(K)>\nonumber\\& &+ \sum_{P\neq K}\sum_{\lambda=1}^{2}\mathcal A^{(\alpha)}_{K}(P,\lambda)|P\otimes\zeta_{\lambda}(P)>]
\label {scstate}
\end{eqnarray}
The new eigenstate $|\Phi^{(\alpha)}(K)>$ is given in terms of the amplitude $\mathcal A^{(\alpha)}_{K}(P,\lambda)$ which represents the overlap between the eigenstate $|\Phi^{\alpha}(K)>$ and the spinor, $|P\otimes\zeta_{\lambda}(P)>$. The amplitude  $|\Phi^{(\alpha)}(K)>$is computed with the help of the matrix elements of the scattering potential given in eq.$26$ .The  function ${N^{(\alpha)}(K)}$ represents the normalization factor defined by the orthonormality condition, $<\Phi^{(\alpha)}(K)|\Phi^{(\beta)}(P)>=\delta_{\alpha,\beta}\delta_{\vec{K}}{\vec{P}}$.
The non-periodic potential is time reversal invariant therefore the zero momentum eigenstate $|\Phi^{(\alpha)}(K=0)>$ remains double degenerated!
In order to deal with the state $\vec{K}=0$ we replace the integrals over the momenta with a discrete sum and replace the two dimensional delta function $\delta^{2}(\vec{K}=0)$ with the Kronecker delta function $\delta_{\vec{K},0}$.
We will use the $new$ $basis$ given in eq. $26$ to  represent the coordinates (similar to eq.4). We find the matrix elements of the Cartesian commutator in the new basis (the basis due to the scattering potential).

\begin{equation}
<\Phi^{(\alpha)}(K)|[r^{(1)},r^{(2)}] |\Phi^{(\beta)}(P)>=[-i\frac{1}{2}(\sigma^{(1)})_{\alpha,\beta}]\delta_{\vec{K},\vec{P}}\delta_{\vec{K},0}[{N^{(\alpha)}(0)}{N^{(\beta)}(0)}]^{-\frac{1}{2}}
\label{comscatter}
\end{equation}
Using the result  given in eq.$27$ we compute the spin Hall current following the steps  given in eq.$11$. The $static$ current is determined by the normalization function $N^{(\alpha)}(\vec{K}=0)$ given by  eq.$26$.

\begin{equation} J^{(3,ext.static)}_{2}=\frac{-e}{2}E^{(ext)}_{1}\frac{1}{2}\sum_{\alpha=1}^{2}\sum_{K}\delta_{\vec{K},0}[\frac{1}{N^{(\alpha)}(K)2\pi}]=[\frac{1}{2}\sum_{\alpha=1}^{2}\frac{1}{N^{(\alpha)}(\vec{K}=0)}]\frac{-e}{4\pi}E^{(ext)}_{1}
\label{result}
\end{equation}

 The normalization factor $N^{(\alpha)}(\vec{K}=0)$ is a function of the amplitude $\mathcal A^{(\alpha)}_{K=0}(P,\lambda)$ (the projection of the eigenstate $|\Phi^{(\alpha)}(K=0)>$ on the spinor $|P\otimes\zeta_{\lambda}(P)>$ ). From eq.$26$ we determine he normalization factor,
 \begin{equation}
  \frac{1}{2}[\sum_{\alpha=1}^{2}\frac{1}{N^{(\alpha)}(\vec{K}=0)}]=\frac{1}{2}\sum_{\alpha=1}^{2}[1+\sum_{\lambda=1}^{2}\int\frac{d^{2}P}{(2\pi)^{2}}|\mathcal A^{(\alpha)}_{K=0}(P,\lambda)|^{2}]^{-1}
\end{equation}  
   The amplitude $\mathcal A^{(\alpha)}_{K=0}(P,\lambda)$ is a function of the matrix elements $V(\vec{q})$ and is obtained within perturbation theory ,  
    $|\mathcal A^{(\alpha)}_{K=0}(P,\alpha)|^{2}=[\frac{V(\vec{K=0}-\vec{P})}{(\frac{(\hbar^{2}}{m})^{2}[k_{SO}|P|+\frac{1}{2}P^{2}]}]^{2}$.
     This allows to compute the current in eq.$28$. We substitute the amplitude $A^{(\alpha)}_{K=0}(P,\lambda)$ into the  normalization function
$N^{(\alpha)}(\vec{K}=0)$ and obtain from eq.$28$ the current;
\begin{equation} J^{(3,ext.-static)}_{2}=[1+\frac{m/\hbar^{2}}{(\hbar^{2}/2m)k^{2}_{so}}\frac{1}{2\pi}\int_{\frac{K_{min}}{k_{so}}}^{\frac{\Lambda}{k_{so}}}\frac{(V(q))^{2}}{q(1+\frac{1}{2}q^{2})^{2}}\,d q]^{-1}[\frac{-e}{4\pi}E^{(ext)}_{1}] 
\end{equation}

In eq.$30$  we have introduced the dimensionless momentum $q$ (the momentum has been normalized  by the spin orbit momentum $k_{so}$). The last integral is evaluated with the help of the dimensionless  infrared cutoff $\frac{K_{min}}{k_{so}}=\frac{\lambda_{so}}{L}$ where $L$ is the size of the two dimensional system, $\lambda_{so}$ is the spin orbit wavelength and $\Lambda$ is an arbitrary ultraviolet cutoff (which does not affect the integration). The momentum integration can be performed when the matrix elements $(V(q))^{2}$ are replaced by a momentum independent potential. This introduces the scattering strength  exponent $\gamma_{sc}$ ,$\gamma_{sc}\equiv\frac{(V(q))^{2} m/\hbar^{2}}{\hbar^{2}/m k^{2}_{so}}$. 

We observe that for an infinite systems the integral diverges causing the   spin Hall current to $decreases$ with the $size$ of the system $L$,

$J^{(3,ext.-static)}_{2}=[1+ \gamma_{sc}Ln(\frac{L}{\lambda_{so}})]^{-1}[\frac{|e|}{4\pi}E^{(ext)}_{1}]$

Next we add the time dependent part spin Hall current.This is done using equations $10-15$.In the absence of inelastic scattering   $\Omega_{o}\tau_{s}\rightarrow\infty$  we find following the result in eq.15 that  the  the $total$ spin Hall current  is given by,
\begin{equation}
J^{(3)}_{2}= [1+ \gamma_{sc}Ln(\frac{L}{\lambda_{so}})]^{-1}[\frac{|e|}{8\pi}E^{(ext)}_{1}]
\end{equation}

We observe that for an infinite systems the integral diverges causing the   spin Hall  current to $decreases$ with the $size$ of the system $L$.
The spin Hall current vanishes for an infinite system since the scattering causes the weight of the zero momentum state to spread over to all other states. Since the integrated weight for the non zero states diverges in two dimensions we find that the projection of the eigenstate with the  $K=0$ state vanishes. This result has been obtained using general arguments, the only assumption is that the absolute value of the square of the scattering potential is momentum independent, such a situation is realized for a $single$ $impurity$ potential.
For many impurities we obtain the same result as for the single impurity if we replace the square of the scattering potential by an ensemble average over the impurity configuration, $V(\vec{q}))^{2}\approx<(V(\vec{q}))^{2}>_{configuration-average}=V_{sc}^{2}$ and obtain the scattering exponent $\gamma_{sc}\equiv\frac{V_{sc}^{2} m/\hbar^{2}}{\hbar^{2}/m k^{2}_{so}}$.  For $many$ $impurities$  the fluctuations of the scattering potential  will cause   the current  to vanishes  faster than our power law prediction.

\section{CONCLUSION}

 A new formulation for the spin Hall current  based on exact ground state of the Rashba Hamiltonian  has been presented. This formulation shows that  the degeneracy of the wave function at $\vec{K}=0$ gives rise to non-commuting coordinates. Due to the fact that the spin Orbit wave function has no gap the spin Hall conductivity is composed from two contribution ,a static part caused by the non-commuting coordinates and a time dependent linear response part.
When the Zeeman interaction is larger than the polarization energy 
the spin Hall conductivity is determined by the static part.
The uniform magnetization is computed and we  show that the spin current which is covariantly conserved  is effectively conserved to first order in the electric field.

Using this new formulation we find the $exact$ value for the spin Hall conductivity which   vary  from $\frac{|e|}{4\pi}$ to $\frac{|e|}{8\pi}$.

 An exact derivation for the vanishing the spin Hall current caused by a scattering potentials has been presented. The spin Hall  current vanishes for an infinite system since the scattering causes the weight of the zero momentum state to spread over to all other states. Since the integrated weight for the non zero states diverges in two dimensions we find that the projection of the eigenstate with the  $K=0$ state is zero. This result has been obtained using general arguments, the only assumption is that the absolute value of the square of the scattering potential is momentum independent, such a situation is realized for a $single$ $impurity$ potential.

\appendix 
\section{THE SU(2) BERRY PHASE}

  The effect of the static electric field is equivalent to a time dependent vector potential . As a result obtain a   time dependent momentum  vector, $K^{(1)}(t)=K^{(1)}(0)+\frac{e}{\hbar}E^{(ext)}_{1}t$, $K^{(2)}(t)=K^{(2)}(0)$ .Therefore according to the adiabatic approximation we  replace
the spinors $|K\otimes\zeta_{\alpha}(K)>$ by the time dependent  state,$|K(t)\otimes\zeta_{\alpha}(K(t))>$.
According to Berry  \cite {Berry,barry} one finds instead of the spinor $|K(t)\otimes\zeta_{\alpha}(K(t))>=exp[\frac{-i}{\hbar}\int_{0}^{t}\epsilon_{\sigma(\alpha)}(K(t^{'}))\,d t^{'}]|K(0)\otimes\zeta_{\alpha}(K(0))>$ the $SU(2)$ time dependent solution $|\Psi_{\alpha}(K,t)>$  is given by, 

\begin{equation} 
 |\Psi_{\alpha}(K,t)>=T exp[-i\int_{0}^{t}\Omega(K(t^{'})\,d t^{'}]exp[\frac{-i}{\hbar}\int_{0}^{t}\epsilon_{\sigma(\alpha)}(K(t^{'}))\,d
 t^{'}]|K(0)\otimes\zeta_{\alpha}(K(0))>
 \label{Berry}
 \end{equation}
 
 We observe that the Berry phase has been replaced by a time ordered matrix, $T exp[-i\int_{0}^{t}\Omega(K(t^{'})\,d t^{'}]$ with the $SU(2)$ matrix elements, $\Omega_{\alpha,\alpha}(K(t))=<K(t)\otimes\zeta_{\alpha}(K(t))|
 \frac{d}{d t}|K(t)\otimes\zeta_{\alpha}(K(t))>$ and non-diagonal matrix elements,
   $\Omega_{\alpha=1,\beta=2}(K(t))=exp[\frac{-i}{\hbar}\int_{0}^{t}(\epsilon_{\sigma(\alpha=1)}(K(t^{'}))-\epsilon_{\sigma(\beta=2)}(K(t^{'})))\,d t^{'}]<K(t)\otimes\zeta_{\alpha=1}(K(t))|
 \frac{d}{d t}|K(t)\otimes\zeta_{\beta=2}(K(t))>$.
 The $SU(2)$ spinor $|\Psi_{\alpha}(K,t)>$ is used to construct the time dependent representation for the coordinate and the velocity operator.

\begin{eqnarray} \nonumber
 &&r^{(i)}(t)=\int\frac{d^{2}K}{(2\pi)^{2}}\int\frac{d^{2}P}{(2\pi)^{2}}\sum_{\alpha=1}^{2}\sum_{\beta=1}^{2}|K\otimes\zeta_{\alpha}(K)><\zeta_{\beta}(P)\otimes P| \\
& &T (e^{i\int_{0}^{t}\Omega(K(t^{'}))\,d t^{'}})e^{\frac{i}{\hbar}\int_{0}^{t}\epsilon_{\alpha}(K(t^{'}))\,d
 t^{'}}r^{(i)}_{\alpha,\beta}(K,P)e^{\frac{-i}{\hbar}\int_{0}^{t}\epsilon_{\beta}(K(t^{'}))\,d
 t^{'}}T( e^{-i\int_{0}^{t}\Omega(K(t^{'})\,d t^{'}})
\label{bdynamics}
\end{eqnarray}
where $r^{(i)}_{\alpha,\beta}(K,P)$ is the matrix element given in eq.4. The time derivative of the coordinate representation given in eq.$A 1$ represents the velocity operator.

\begin{eqnarray}\nonumber
\frac{d r^{(2)}(K(t))}{d t}&=&\frac{1}{i\hbar}[r^{(2)}(K(t)),h_{0}(K(t))]+\frac{1}{i\hbar}[r^{(2)}(K(t)),\Omega(K(t))]\nonumber \\& &=\frac{1}{i\hbar}[r^{(2)}(K(t)),h_{0}(K(t))]+\frac{1}{i\hbar}(-e) E^{(ext)}_{1}[r^{(2)}(K(t)),r^{(1)}(K(t)) ]
\label {bvelocity}
\end{eqnarray}
 The first term in eq.$A2$ represents the velocity in the absence of the external field. The second term describes the effect of the external field proportional to the commutator of the Cartesian coordinates. This can be seen in the following way, the matrix $\Omega(K(t))$ is equal to the product between the time derivative of the momentum and a new matrix $\hat{\Omega}(\frac{d}{d K^{(1)}})$. For this matrix we replace the  momentum derivative by the time derivative momentum derivative. We find, $\Omega(K(t))=\frac{d K^{(1)}}{d t}\hat{\Omega}(\frac{d}{d K^{(1)}})$. The momentum time derivative is proportional to the electric field and
$the$ $matrix$ $\frac{d}{d K^{(1)}}$ is $identified$ with the $Cartesian$ $coordinate$ ,$[r^{(2)},\hat{\Omega}(\frac{d}{d K^{(1)}})]\equiv[r^{(2)},r^{(1)}]$. As a result of the $SU(2)$ adiabatic evolution , the velocity operator given in eq.$A2$  is equivalent to the the Heisenberg equation of motion given in eq.6 (with the same commutation rules as given in eq.4).

\section{GAUGE INVARIANCE IN THE MOMENTUM SPACE-THE COVARIANTLY CONSERVED SPIN CURRENT}
 
In this appendix  we will the consider the  $U(1)\times SU(2)_{(spin)}$ gauge invariance of Ahharon-Casher (A-C)\cite {Casher} model for a periodic lattice  in the momentum space.  
 The A-C effect is viewed as  a moving magnetic moment which is equivalent to a moving magnetic current, due to the Lorentz transformation an induced electric-charge appears in the laboratory frame, which interacts with an electrostatic potential.
 
\begin{equation}
H = \frac{\hbar^2}{2m } [ - i \vec{\partial} - \frac{e}{\hbar c} \vec{A}(r) - \frac{(g-1) \mu_B}{2 \hbar c} (\vec{\sigma}\times \vec{E}(r) ) ]^2 - \frac{g}{2} \mu_B \vec{\sigma} \cdot (\vec{\nabla} \times \vec{A}) 
\end{equation}

where $\vec{A}(r)$ is the $U(1)$ electromagnetic vector potential, $A_E(r) \equiv (\vec{\sigma} \times \vec{E} (r) ) $ is the $SU(2)$ ``electrostatic vector potential'' ($\vec{\sigma}$ is the Pauli matrix and $\vec{E}(r)$ is the electrostatic field), $\mu_B$ is the Bohr magneton, $g \approx 2$ is the gyro magnetic factor with $g\rightarrow(g-1)/2$ being the Thomas precession and  ``$e$'' is the electrostatic charge.

For a periodic lattice we use the quasi-momentum representation with momentum integration restricted to the first Brillouin zone.  The single particle energy and the two component spinor obey the  symmetry in the momentum space (for simplicity we replace  the Bloch wave function by the  free particle representation),   $\epsilon(\vec{K}) = \epsilon(\vec{K} + \vec{G})$ , $\Psi^+(\vec{K}) = \Psi^+(\vec{K} + \vec{G})$, $\Psi(\vec{K} ) = \Psi(\vec{K} + \vec{G})$ where $\vec{G}$ is the reciprocal Lattice vector.
 In order to describe the  two dimensional Rashba model we take  the $z$ component of the electric field to be constant  in space . We introduce the static SU(2)  Rashba term,
 
 $\sum_{A=1,2} \vec{\omega}_A(\vec{K}) \sigma^{(A)} =k_{so} (\vec{\sigma} \times \vec{e}_3)$
  
 Only two components are nonzero; ${\omega}_{(1,A=2)}(\vec{K}) \sigma^{(A=2)}=-k_{so}\sigma^{(2)}$, 
  ${\omega}_{(1,A=1)}(\vec{K}) \sigma^{(A=1)}=k_{so}\sigma^{(1)}$  and ${\omega}_{(1,A=1)}(\vec{K}) \sigma^{(A=1)}= {\omega}_{(2,A=2)}(\vec{K}) \sigma^{(A=2)}=0$. The fluctuation of   the electric field gives rise to the $SU(2)$  gauge field, $\vec{\Lambda}(\vec{K}) = \sum_{A=1,2} \vec{\Lambda}_A (\vec{K})\sigma^{(A)}$ .In the absence of a magnetic field we have $\sum_{A = 1,2,3} \omega_{0,A} (\vec{K}) \sigma^{(A)}=0$ and the fluctuating part is given by ,$\sum_{A = 1,2,3} \Lambda_{0, A} (\vec{K}) \sigma^{(A)}$. 
  In the momentum representation the model given in eq.$B1$ the external gauge  field and  $without$ the $static$ $magnetic$ $field$   takes form :
 $S = S_0 + S_{ext} $ , $S_0$ is the free electron action in the presence of the Rashba term  and $S_{ext}$ is the electromagnetic 
 $U_{em}(1) \times SU(2)_{spin}$ action.
\begin{eqnarray}
S_0 &=& \int dt \int \frac{d^{2}K}{(2\pi )^{2}}\{ \Psi^+(\vec{K}, t) ( -i \hbar \partial_t - E_F ) \Psi(\vec{K},t) \nonumber\\&-&
\Psi^+(\vec{K}, t) [ \epsilon(\vec{K} - \sum_{A = 1,2} \vec{\omega}_A (\vec{K}) \sigma^{(A)}) + \sum_{A = 1,2,3} \omega_{0,A} (\vec{K}) \sigma^{(A)} ] \Psi(\vec{K},t) \}
\label{momentum action}
\end{eqnarray}

 In momentum space the  external vector potential $\vec{A} (\vec{q}, t)$ and the scalar potential $A_0(q,t)$  generate trough the A-C term  a  vector \textit{$SU(2)$ gauge potential}  $\vec{\Lambda}(q) = \sum_{A=1,2} \vec{\Lambda}_A(q) \sigma^{(A)}$  and a static \textit{$SU(2)$ gauge potential}  $\sum_{A = 1,2,3} \Lambda_{0, A} (q) \sigma^{(A)}$ .
  The external action \cite{Schri} is given by; $S_{ext}= S_{ext}^{(para)} + S_{ext}^{(dia)}$:
 
\begin{eqnarray} 
S_{ext}^{(para)} &=& \int dt \int \frac{d^{2} q}{(2\pi)^{2}}  \int\frac{d^{2} K}{(2\pi)^{2}}  \Psi^+(K) [ \frac{\hbar}{m} (\vec{K} - \frac{1}{2} \vec{q} - \sum_A \vec{\omega}_A \sigma^{(A)}) \cdot (e \vec{A} (-\vec{q}) + \sum_{A=1,2} \vec{\Lambda}_A(-\vec{q}) \sigma^{(A)})\nonumber \\ 
&+& e A_0 ( - \vec{q}) + \sum_{A=1,2,3} \Lambda_{0,A} (-\vec{q}) \sigma^A] e^{i \vec{q} \cdot \vec{R}(\vec{K})} \Psi(\vec{K}) 
\end{eqnarray} 
\begin{equation}
S_{ext}^{(dia)} = \int dt \int \frac{d^{2} q}{(2\pi)^{2}}  \int \frac{d^{2} K}{(2\pi)^{2}} \frac{\hbar^2}{2 m} \ \Psi^+(\vec{K},t) [ e \vec{A} (\vec{q}) + \sum_{A=1,2} \vec{\Lambda}_{A} (\vec{q}) \sigma^A] \cdot [ e\vec{A}(-\vec{q}) + \sum_A \vec{\Lambda}_A(-\vec{q}) \sigma^{(A)} ] e^{i \vec{q} \cdot \vec{R}(\vec{K})} \Psi(\vec{K},t)
\end{equation}

We observe that the action is $U(1) \times SU(2)$ gauge invariant in momentum space. As a result  the $coordinate$ is  $SU(2)$ $transformed$ , $\vec{R} (\vec{K}) \rightarrow \vec{r}(\vec{K})$. Using Noether's  theorem \cite{Weinberg} we obtain that the $Spin$ - $current$  is determined  by  the derivative with respect to the $SU(2)$ gauge potential $\Lambda_{\mu,A}(\vec{q})$.
\begin{equation}
\hat{J}_0^{(A)}(\vec{q}) = \frac{\hbar}{2} \frac{\partial S}{\partial \Lambda_{0,A}(-q)} = \frac{\hbar}{2} \int \frac{d^{2} K}{(2\pi)^{2}} \Psi^+(K) \sigma^{(A)} e^{i \vec{q} \cdot \vec{R}} \Psi(\vec{K})
\label{spin-density}
\end{equation}

\begin{equation}
\hat{J}_i^{(A)}(\vec{q}) = \frac{\hbar}{2} \frac{\partial S}{\partial \Lambda_{i,A}(-q)} = \frac{\hbar}{2} \int \frac{d^{2} K}{(2\pi)^{2}}\Psi^+(K) [\frac{1}{2}\{ \frac{\hbar}{m}(\vec{K} - \frac{1}{2}\vec{q} - e \vec{A}(q) )_{i}, \sigma^{(A)} \}_{+}    ] e^{i \vec{q} \cdot \vec{R}(\vec{K})} \Psi(\vec{K})
\label{spin-current}
\end{equation}
where $\sigma^{(A)}$ are the Pauli matrices.

In order to study the electromagnetic response for an  electric  field  without an orbital magnetic effect $\vec{q}\times\vec{A}(\vec{q})=0$ we take the vector potential to be zero    $\vec{A}(\vec{q})=0$ and represent  the electric field by a  $scalar$ $potential$ $A_{(0)}^{(ext)} (  \vec{q})$ which satisfies  the condition $E_{(1)}^{(ext)}(\vec{q}, t) = i q_{(1)} A_0^{(ext)}(\vec{q},t)$. (An alternative choice is to replace the vector potential by a time dependent vector potential. As a result we have to use a $time$ $dependent$ $momentum$ $K^{(1)}(t)=K^{(1)}(0)+\frac{e}{\hbar}E_{1}^{(ext)}t$ this approach has been used in the appendix $A$.  )
From eq.$B6$ we obtain that the spin  velocity is given by the symmetric product between the spin operator and the velocity operator. Using the SU(2) transformation which diagonalizes the Rashba Hamiltonian introduced in chapter $III$ with the velocity operator 
$V_{i}(\vec{K})=(\frac{d r^{(i)}(\vec{K},t)}{dt})|_{t=0}$ gives us according to eq.$B6$ the  spin velocity operator , ,$\ell^{(A)}_{i}(\vec{K})=\frac{1}{2}\{V_{i}(\vec{K}),\hat{\sigma}^{(A)}\}_{+}$ where 
 $\{,\}_{+}$ stands for the symmetric product and $\hat{\sigma}^{(A)}=U^{\dagger}(\vec{K})\sigma^{(A)}U (\vec{K})$ represents the transformed Pauli matrix with $A=1,2,3$.
 The \textit{transformed spin current operator in the Heisenberg picture } is given by;
\begin{equation} 
  J_{i,H}^{(A)}(\vec{q},t)  = (\frac{\hbar}{2}) \int\frac{d^{2} K}{(2\pi)^{2}} \Psi^{+}(\vec{K}) [ \frac{1}{2}\{  {V}_{i,H}(\vec{K}), \hat{\sigma}_{H}^{(A)}(\vec{K},t)\} _{+} e^{i \vec{q} \cdot \vec{r}_{H}(\vec{K},t)} - \frac{\hbar}{m} (\frac{1}{2} \vec{q})_{i} \{ e^{i \vec{q} \cdot \vec{r}_{H}(\vec{K},t)} , \hat{\sigma}_{H}^{(A)}(\vec{K},t) \} _+ ] {\Psi}(\vec{K}) 
\label{transformed current}
\end{equation}
 
where $\hat{\sigma}^{(A)} = U^+(K) \sigma^{(A)} U(K)$ with  the Heisenberg representation of the  matrix elements, ${V}_{i,H}(\vec{K},t)$, $r^{(i)}_{H}(\vec{K},t)$ and 
 $\hat{\sigma}_{H}^{(A)}(\vec{K},t)$.  
The spin currents are computed by taking the expectation value with respect the Many-Body state,    $|F.S.>=|K_{F}^{\uparrow},K_{F}^{\downarrow}>$ ,

\begin{equation}
J_{i}^{(A)}(\vec{q},t)=<F.S.|J_{i,H}^{(A)}(\vec{q},t)|F.S.>
\end{equation}

 Using  Noether's theorem \cite{Weinberg,Frohlich}, we obtain that the spin currents are only \textit{covariantly conserved}.
\begin{equation}
\partial_t J_0^{(A)}(\vec{q},t) + i \vec{q} \cdot \vec{J}^{(A)}(\vec{q},t) = 2 \epsilon_{ABC} [ (\omega_{0,B} + \Lambda_{0, B}(-\vec{q},t)) J_0^{(C)}(\vec{q},t) + (\vec{\omega}_B + \vec{\Lambda}_B(-q))\cdot \vec{J}^{(C)} (\vec{q},t) ]
\label{continuity}
\end{equation}
 where $\epsilon_{ABC}$is the antisymmetric tensor which takes the values,$1$,$0$,$-1$.
 We apply this equation to the $Rashba$ model.
  The spin current components in the limit $q \rightarrow 0$ obey the following  continuity equation for the 
 three polarizations:
  $\partial_t J_0^{(1)} + i \vec{q} \cdot \vec{J}^{(1)}(\vec{q}) = 0$, $\partial_t J_0^{(2)} + i \vec{q} \cdot \vec{J}^{(2)}(q) = 0$. Only the $A=3$ component \textit{violates the continuity equation}
\begin{eqnarray}
& &\partial_t J_0^{(3)} + i \vec{q} \cdot \vec{J}^{(3)}(\vec{q}) = 2 \epsilon_{3BC} \vec{\omega}_B(\vec{K}) \cdot \vec{J}^{(C)}(q) = 2 k_{so} (J_1^{(1)}(q) + J_2^{(2)}(q))\nonumber\\&_{\overrightarrow{q \rightarrow0}}&2 k_{so}(\frac{\hbar}{2}) <F.S.| \int \frac{d^2 K}{(2\pi)^{2}} \Psi_{H}^+(\vec{K},t) [ \frac{1}{2}\{{V}_1(\vec{K}),\hat{ \sigma}^{(1)}\}_{+} + \frac{1}{2}\{{V}_2(\vec{K}),\hat{ \sigma}^{(2)}\}_{+} ] \Psi_{H}(\vec{K},t)|F.S.>\nonumber\\&=&2 k_{so}(\frac{\hbar}{2}) \int \frac{d^2 K}{(2\pi)^{2}}<F.S.|\Psi^+(\vec{K})[\frac{1}{2}\{\sigma_{H}^{(3)}(\vec{K},t),(\vec{V}_{H}(\vec{K},t)\times \frac{\vec{K}}{|\vec{K}|})\}_{+}\nonumber\\&+&\frac{1}{2}\{\sigma_{H}^{(2)}(\vec{K},t),(\vec{V}_{H}(\vec{K},t)\cdot \frac{\vec{K}}{|\vec{K}|})\}_{+}]\Psi(\vec{K})|F.S.>
\label{covariant}
\end{eqnarray}
The first term on the right hand side of  the equation $ $ vanishes  as a result  of to the momentum  integration .The second term vanishes as a result of the expectation value with respect the Pauli matrices. The  external electric field induces 
a  first order contribution to the spin velocity  which  vanishes after the momentum integration is performed.Therefore  we conclude that the  $continuity$ $equation$ is satisfied effectively.

\section{THE SPIN  AND CHARGE HALL EFFECT IN THE PRESENCE OF A ZEEMAN INTERACTION}

In this appendix  we will consider  the   charge Hall effect in the $static$ linear response approximation.
We will consider the  current for the following cases :
 No orbital magnetic  contribution, magnetic orbital current without spin Orbit and at the end we compute the charge Hall in the presence of  the spin Orbit interaction and a weak magnetic field.
 
 In the $absence$ of the orbital motion the charge Hall conductivity  is given by; 
\begin{equation}
\frac{e^{2}}{2h}\left[f_{F.D.}(\epsilon_{so}+b_{3}-E_{F})-f_{F.D.}(-b_{3}+\epsilon_{so}-E_{F})\right]
\end{equation}
 This result follows directly from equation $5$ once we replace the Pauli matrix $\sigma^{(3)}$ with the identity matrix $I$.  The conductivity  is zero in the absence of the Zeeman  interaction and vanishes at zero temperature. 

-The orbital effects are investigated for  a periodic potential $W(\vec{q})$ with the Bloch eigenfunctions $u_{n,K}(\vec{q})$.  $n$ is the band index and $\vec{q}$ is the coordinate in the unit cell. The coordinate $\vec{R}$ is a function of  $\vec{q}$, $\vec{R} = \vec{q} + \vec{r} \equiv \vec{q} + i \frac{\vec{\partial}}{\partial K}$. The curvature is determined by the band connection  $\vec{A}_{n,m}(\vec{K}) = i \int \frac{d ^{2}q }{(2\pi)^{2}}u_{m,K}^{\ast} (\vec{q}) \frac{\vec{\partial}}{\partial K} u_{n,K}(\vec{q})$. In the one band approximation the commutator is given by, $\left[r^{(1)},r^{(2)}\right]=i\Omega(\vec{K},n)$ with the band curvature, $\Omega(\vec{K},n)=\frac{\vec{\partial}}{\partial K} \times \vec{A}_{n,n}(\vec{K})$.

We apply this formalism to electrons in a constant magnetic field $b$ perpendicular to the two dimensional plane. (For a strong magnetic field which is such that the magnetic flux per unit cell is a rational multiple $\frac{p}{l}$ of the the flux unit $\Phi_{0}$ we obtain a magnetic unit cell with $l$ cites and a reduced Brillouin zone.  The   complex eigenfunction $u_{n,K}(\vec{q})$ has zero's in the Brillouin  zone which is the origin for the $U(1)$ Berry phase connections.)  
In order to study orbital effects we replace the  Rashba Hamiltonian  equation  by, $
h_{0}(\vec{K}, \vec{r}) = \frac{\hbar^2}{2m}(\vec{K} - \frac{e}{\hbar c} \vec{b} \times  \vec{r} - k_{so}(\vec{\sigma}\times \hat{e}_3))^2 $.   We define the $kinetic$ $crystal$ $momentum$ $\vec{\kappa}$ which replaces the momentum $\vec{K}$  by, $\vec{\kappa} = \vec{K} - \frac{e}{2 \hbar c} \vec{b}\times \vec{r}$. The $kinetic$ $crystal$ $momentum$ $components$ do $not$ $commute$ in a magnetic field, $[\kappa_1, \kappa_2] = - i \frac{e}{\hbar c} b$. 

In the $absence$ of the $spin-orbit$ interaction the Bloch function in a magnetic field \cite{nij}  gives rise to  a U(1)
Berry phase connection $\vec{A}_{n,n}(\vec{K}) = i \int \frac{d ^{2}q }{(2\pi)^{2}}u_{n,K}^{\ast} (\vec{q}) \frac{\vec{\partial}}{\partial K} u_{n,K}(\vec{q})$  \cite{qian} and a Berry curvature $\Omega(\vec{K},n)=\frac{\vec{\partial}}{\partial K} \times \vec{A}_{n,n}(\vec{K})$.   In the one band approximation we find $\left[r^{(1)},r^{(2)}\right]=i\Omega(\vec{K},n)$. We introduce the $4\times4$  Symplectic matrix  to describe the  commutators of the Cartesian coordinates and   kinetic momentum,  $J_{i,j} = [ \xi_i, \xi_j]$ where $\vec{\xi} = (r_1, r_2, \kappa_1, \kappa_2) \equiv (\xi_1, \xi_2, \xi_3, \xi_4)$. The Heisenberg equation of motion  are given by,
\begin{equation}
 (i\hbar) \frac{d \xi_i(\vec{K})}{d t} = \sum_j J_{i,j} \frac{\partial h_{0}(\vec{K}, \vec{r}(\vec{K}))}{\partial \xi_j(\vec{K})}
 \end{equation}
 We obtain the same equation of motion as the one given in ref.\cite {haldane} (see eqs.$1,2$). Using the equation of motion we compute the charge Hall current in the $i=2$ direction,
\begin{equation}
 J_{2, Hall}^{(c)} = \frac{e}{L^2} \sum_{\vec{K},n} Tr [ \frac{d \hat{r}_2(\vec{K})}{d t} F_{F.D.}(E_{n}(\vec{K}), E_F) ]
\end{equation} 
   ``Tr'' stands for the trace over the spin degrees of freedom. $F_{F.D.}(\vec{K}, E_F)=\delta(\sigma,\uparrow)	f_{F.D.}^\uparrow(E_{n}(\vec{K}), E_F)+ \delta(\sigma,\downarrow) f_{F.D.}^\downarrow(E_{n}(\vec{K}), E_F)$ is the Fermi-Dirac function and  $E_{n}(\vec{K})$  are the Bloch eigenvalues as a  function of the  magnetic band index $n$. We find that the conductivity is given by,
   
\begin{equation}
\frac{e^{2}}{h}\int d^{2}K\sum_{n} \Omega(\vec{K},n)[ f_{F.D} (E_{n}(\vec{K})+b_{3}-E_{F} ) + f_{F.D}(E_{n}(\vec{K})-b_{3}-E_{F})] 
\end{equation}

This result is in agreement with the result in ref.\cite{haldane} ( see eq.$3$). For strong magnetic fields the commutator  $\left[r^{(1)},r^{(2)}\right]=i\Omega(\vec{K})$  describes  the vorticity  for the Hall wave function \cite{kohm}.  We find that the conductance is determined by the integral over the curvature and is given by the first Chern number \cite{kohm} (see eqs.3.9 and 4.8).

Next we consider the combined effect of a  spin Orbit  interaction   and a $weak$ $magnetic$ field. For each Bloch band with magnetic eigenfunctions   $u_{n,K}(\vec{q})$  we compute  the following $SU(2)$  matrix elements,
 $(\epsilon(\kappa))_{n,n} + (\vec{\sigma}\cdot \vec{\Re})_{n,n}$  where $(\vec{\Re})_{n,n} = [(- \frac{\hbar^2 k_{SO}}{m^*} \kappa_2, \frac{\hbar^2 k_{SO}}{m^*} \kappa_1, \frac{1}{2} g \mu_B(B+ b))]_{n,n} $ is a fictitious magnetic field for the band  $n$ and  $(\epsilon(\vec{\kappa}))_{n,n}\equiv E_{n}(\kappa)$  is the eigenvalue of the Bloch state  and $b_{3}\equiv \frac{1}{2} g \mu_B(B+ b)$ is the total Zeeman field. We introduce an SU(2) transformation to rotate the Pauli matrix $\vec{\sigma}$ in the direction of the fictitious magnetic field $(\vec{\Re})_{n,n}$ where $(\vec{\kappa})_{n,n} = \vec{K} - \frac{e}{2 \hbar c} \vec{b}\times (\vec{r})_{n,n}= \vec{K} - \frac{e}{2 \hbar c} \vec{b}\times\vec{A}_{n,n}(\vec{K})$.   For weak magnetic  field   we find that the curvature for the band  $n$ is given by, $\left[r^{(1)},r^{(2)}\right]_{n,n}=i I\cdot\Omega(\vec{K},n)+\frac{i}{2}\left[-\sigma^{1}\sin\vartheta(\vec{\kappa_{n,n}})+\sigma^{3}\cos\vartheta(\vec{\kappa}_{n,n})\right]2\pi\delta^{2}((\kappa)_{n,n})$. The first term is  the Berry curvature of the  magnetic Bloch function (generated by the connection) $\vec{A}_{n,n}(\vec{K}) = i \int \frac{d ^{2}q }{(2\pi)^{2}}u_{n,K}^{\ast} (\vec{q}) \frac{\vec{\partial}}{\partial K} u_{n,K}(\vec{q})$ \cite {qian}. The second part is the curvature matrix due to the $SU(2)$ spin-orbit interaction. This part  is obtained by replacing $\vec{K}\rightarrow (\vec{\kappa})_{n,n} = \vec{K} - \frac{e}{2 \hbar c} \vec{b}\times (\vec{r})_{n,n}= \vec{K} - \frac{e}{2 \hbar c} \vec{b}\times\vec{A}_{n,n}(\vec{K})$.  As a result we find that the Hall conductivity has two parts, a $ band$ contribution and a $spin$ $Orbit$ part;
\begin{eqnarray}
 & & \frac{e^{2}}{h}\int d^2K\sum_{n}\Omega(\vec{K},n) [ f_{F.D} (E_{n}(\vec{K}) +\epsilon_{so}+b_{3}-E_{F} ) + f_{F.D}(E_{n}(\vec{K})+\epsilon_{so}-b_{3}-E_{F})]\nonumber\\&+&\frac{e^{2}}{2h}\sum_{n}[ f_{F.D} (E_{n}(0)+\epsilon_{so}+ b_{3}-E_{F} ) - f_{F.D}(E_{n}(0)+\epsilon_{so}-b_{3}-E_{F})]
\end{eqnarray} 
  In the present case the spin orbit contribution to the current (the second term)  is determined by the  magnetic cyclotron eigenvalues $E_{n}(0)\approx \frac{e\hbar b}{m c}(n+\frac{1}{2})$, $n=0,1,2,3,...$. The spin Orbit contribution to the current decreases with the increase of the magnetic field faster than in the case where orbital effects are absent.


\begin{thebibliography}{99}
\bibitem{wund}J.Wunderlich,B.Kaestner,J.Sinova,and T.Jungwirth,Phys.Rev.Lett.
\textbf{94},047204(2005).
\bibitem{aw}Y.K.Kato,R.C.Myers,A.C.Gossard and D.D.Awschalom,Science \textbf{306},1910(2004).
\bibitem{Sih} V.Sih,R.C.Myers,Y.K.Kato,W.H.Lau,A.C.Gossard and D.D.Awshalom ,Nature Physics \textbf{1},31-35(2005).
\bibitem{Bern} B.Andrei  Bernevig and Shou-Cheng Zhang  cond-mat/0412550.
\bibitem{hir}J.E.Hirsch,Phys.Rev.Lett.\textbf{83},1834(1999).
\bibitem{sinova}J.Sinova et al.,Phys.Rev.Lett.
\textbf{92},126603(2004).
\bibitem{halperin}E.G.Mishchenko,A.V.Shytov.and B.I.Halperin,Phys.Rev.Lett.93,226602(2004)
\bibitem{sch}D.Schmeltzer,cond-mat/0406565.
\bibitem{culcer}D.Culcer,A.MacDonald and Q.Niu ,Phys.Rev.B\textbf{68},045327(2003).
\bibitem{yang}S.Zhang and Z.Yang,Phys.Rev.Lett.\textbf{94},066602(2005).
\bibitem{mol}J.I.Inoue,G.E.W.Bauer and L.W.Molenkamp,Phys.Rev.B\textbf{70},041303(2004).
\bibitem{los}S.I.Erlingsson,J.Schliemann and D.Loss,Phys.Rev.B\textbf{71},035319(2005).
\bibitem{di}P.Zhang,J.Shi,D.Xiao and Q.Niu ,cond-mat/0503505.
\bibitem{sheng}D.N.Sheng,L.Sheng,Z.Y.Weng and F.D.M.Haldane,cond-mat/0504218.

\bibitem{david}D.Schmeltzer,cond-mat/0504035.



\bibitem{foldy}L.L.Foldy,and S.A.Wouthuysen,Phys.Rev.\textbf{78},29(1950).
\bibitem{winkler}R.Winkler ,Springer Tracts in Modern Physics.\textbf{191}.
\bibitem{Jackson}J.D.Jackson,Classical Electrodynamics ,2nd ed. Wiley New York \textbf(1975).
\bibitem{lutt}J.M.Luttinger,Phys.Rev.\textbf{102},1030(1956).
\bibitem{mura}S.Murakami,N.Nagaosa,S-C.Zhang,Phys.Rev.B\textbf{69},235206(2004).
\bibitem{frankel}T.Frankel,"'The Geometry of Physics"',pages ,156,253-254,422 ,Cambridge Press , 2004.
\bibitem{naka}M.Nakahara, "`Geometry,Topology and Physics"',Adam Hilger,1991.
\bibitem{Casher}Y.Aharonov and A.Casher,Phys.Rev.Lett.\textbf{53},319 (1984). 
\bibitem{kohm}M.Kohmoto,Anals of Physics \textbf{160},343(1985).
\bibitem{zak}J.Zak, Phys.Rev.B \textbf{40},3156(1989).

\bibitem{rash}Yu.A.Bychov and E.I.Rashba,J.Phys.C\textbf{17},6039(1984).

 
\bibitem{Berry}M.V.Berry Proc.R.Soc.Lond.\textbf{A 392},45-47,(1984).
\bibitem{barry} B.Simon Phys.Rev.Lett.\textbf{51},2167(1983).



\bibitem{Schri}J.R.Schrieffer, "`Theory of Superconductivity"',pages 206-212, Addison Wesley (1966) 
\bibitem{Weinberg}S.Weinberg"'The Quantum Theory Of Fields"',vol.1,pages 306-318,Cambridge University Press(1995).
\bibitem{Frohlich}J.Frohlich and U.M.Studer Rev.of Modern Physics \textbf{65},733(1993)
\bibitem{nij}D.J.Thouless,M.Kohmoto,M.P.Nightingale and M.Den Nijs, Phys.Rev.Lett.\textbf{49},405(1982).
\bibitem{qian}Ming-Che Chang and Qian Niu, Phys.Rev.B\textbf{53},7010(1995)
\bibitem{haldane}F.D.M.Haldane,Phys.Rev.Lett.\textbf{93},206602(2004)
\end{thebibliography}
\end{document}